\documentclass[a4paper,USenglish,cleveref, autoref]{lipics-v2019}

\usepackage{float}
\usepackage{hyperref}
\usepackage[utf8]{inputenc}
\usepackage{amsmath}
\usepackage{amssymb}
\usepackage{booktabs}
\usepackage{xspace}
\usepackage{placeins}
\usepackage{todonotes}
\usepackage{esvect}

\usepackage{pifont}

\DeclareMathOperator{\obj}{\operatorname{obj}}

\renewcommand{\epsilon}{\varepsilon}

\newcommand{\ie}{i.\,e.\xspace}
\newcommand{\eg}{e.\,g.\xspace}
\newcommand{\etal}{et al.\xspace}

\title{Advanced Flow-Based Multilevel Hypergraph Partitioning}

\author{Lars Gottesbüren}{Karlsruhe Institute of Technology, Germany}{lars.gottesbueren@kit.edu}{}{}
\author{Michael Hamann}{Karlsruhe Institute of Technology, Germany}{michael.hamann@kit.edu}{}{}
\author{Sebastian Schlag}{Karlsruhe Institute of Technology, Germany}{sebastian.schlag@kit.edu}{}{}
\author{Dorothea Wagner}{Karlsruhe Institute of Technology, Germany}{dorothea.wagner@kit.edu}{}{}
\authorrunning{L. Gottesbüren, M. Hamann, S. Schlag and D. Wagner}

\Copyright{Lars Gottesbüren, Michael Hamann, Sebastian Schlag and Dorothea Wagner}



\supplement{Source code \url{https://github.com/larsgottesbueren/WHFC}}
\funding{This work was supported by the Deutsche Forschungsgemeinschaft (DFG, German Research Foundation) under grants WA654/19-2, WA654/22-2, and SA933/11-1. The authors acknowledge support by the state of Baden-Württemberg through bwHPC.}

\category{}

\relatedversion{}

\nolinenumbers 

\hideLIPIcs  

\EventEditors{Simone Faro and Domenico Cantone}
\EventNoEds{2}
\EventLongTitle{18th International Symposium on Experimental Algorithms (SEA 2020)}
\EventShortTitle{SEA 2020}
\EventAcronym{SEA}
\EventYear{2020}
\EventDate{June 16--18, 2020}
\EventLocation{Catania, Italy}
\EventLogo{}
\SeriesVolume{160}
\ArticleNo{8}

\begin{document}

\maketitle

\begin{abstract}

The balanced hypergraph partitioning problem is to partition a hypergraph into $k$ disjoint blocks of bounded size such that the sum of the number of blocks connected by each hyperedge is minimized.
We present an improvement to the flow-based refinement framework of KaHyPar-MF, the current state-of-the-art multilevel $k$-way hypergraph partitioning algorithm for high-quality solutions.
Our improvement is based on the recently proposed HyperFlowCutter algorithm for computing bipartitions of unweighted hypergraphs by solving a sequence of incremental maximum flow problems.
Since vertices and hyperedges are aggregated during the coarsening phase, refinement algorithms employed in the multilevel setting must be able to handle both weighted hyperedges and weighted vertices -- even if the initial input hypergraph is unweighted.
We therefore enhance HyperFlowCutter to handle weighted instances and propose a technique for computing maximum flows directly on weighted hypergraphs.

We compare the performance of two configurations of our new algorithm with KaHyPar-MF and seven other partitioning algorithms on a comprehensive benchmark set with instances from application areas such as VLSI design, scientific computing, and SAT solving.
Our first configuration, KaHyPar-HFC, computes slightly better solutions than KaHyPar-MF using significantly less running time.
The second configuration, KaHyPar-HFC*, computes solutions of significantly better quality \emph{and} is still slightly faster than KaHyPar-MF.
Furthermore, in terms of solution quality, both configurations also outperform all other competing partitioners.

\end{abstract}

\section{Introduction}

Graphs are a common way to model pairwise relationships (edges) between objects (vertices). 
However, many real-world problems involve more complex interactions~\cite{GPOverviewBook,LRMS}.
Hypergraphs are a generalization of graphs, where each hyperedge can connect an arbitrary number of vertices.
Thus, hypergraphs are well-suited to model such higher-order relationships.

The balanced $k$-way hypergraph partitioning problem (HGP) asks to compute a partition of the vertices
into $k$ disjoint blocks of bounded weight (at most $(1+\epsilon)$ times the average block weight) such 
that few hyperedges are cut, i.e., connect vertices in different blocks~\cite{GPOverviewBook,ak-rdnps-95,pm-hpc-07}. Since 
hyperedges can connect more than two vertices, several notions of cuts exit in the literature~\cite{ak-rdnps-95}. 
In this work, we consider the \emph{connectivity} objective, which aims to minimize $\sum_{e \in E} \omega(e) (\lambda(e) - 1)$, where
$E$ is the set of hyperedges, $\omega(e)$ denotes the weight of hyperedge $e$, and $\lambda(e)$ denotes the number of blocks connected by hyperedge $e$.
Well-known applications of hypergraph partitioning include VLSI design~\cite{ak-rdnps-95}, the parallelization of sparse matrix-vector multiplications~\cite{patoh}, and storage sharding in distributed databases~\cite{KabiljoKPPSAP17}.
We refer to a survey chapter~\cite[Ch. 3]{SchlagHGP} and two survey articles~\cite{ak-rdnps-95, pm-hpc-07} for an extensive overview.

Since HGP is NP-hard~\cite{l-caicl-90}, heuristic algorithms are used in practice. 
The most successful heuristic for computing high-quality solutions is the three-phase \emph{multilevel} paradigm. 
In the \emph{coarsening phase}, multilevel algorithms first successively contract the input hypergraph to obtain a hierarchy of smaller, structurally similar instances.
After applying an \emph{initial partitioning} algorithm to the smallest hypergraph, the contractions are undone and, at each level, refinement algorithms are used to improve the partition induced by the coarser level.

\subparagraph*{Related Work.}
The most well-known and practically relevant multilevel hypergraph partitioners from different application areas are PaToH~\cite{patoh} (scientific computing), hMetis~\cite{kaks-mhpav-99,kk-mkwhp-99} (VLSI design), KaHyPar~\cite{kahypar-mf-19,hs-icshp-17,ahss-ehpa-17,shhmss-hplrb-16} (general purpose, $n$-level), Zoltan~\cite{dbhbc-phpsc-06, scs-rbcmh-19} (scientific computing, parallel), Mondriaan~\cite{vb-atddd-05} (sparse matrices), and Par$k$way~\cite{tk-pmahp-08} (parallel).
Additionally there are MLPart~\cite{ahk-mcp-98} (restricted to bipartitioning), and HYPE~\cite{mmber-hypem-18}, a single-level algorithm that grows $k$ blocks using a neighborhood expansion~\cite{zwltl-gepnh-17} heuristic.

With the exception of KaHyPar-MF~\cite{kahypar-mf-19}, all multilevel HGP algorithms solely employ variations of the well-known Kernighan-Lin~\cite{kl-efppg-70} (KL) or Fiduccia-Mattheyses~\cite{fm-a-82} (FM) heuristics in the refinement phase.
These algorithms repeatedly move vertices between blocks prioritized by the improvement in the objective function.
While they perform well for hypergraphs with small hyperedges, their performance deteriorates in the presence of many large hyperedges~\cite{DBLP:journals/siamsc/UcarA04}.
In this case, many single-vertex moves have no immediate effect on the objective function because the vertices of large hyperedges are likely to be distributed over multiple blocks.
KaHyPar-MF~\cite{kahypar-mf-19} therefore additionally employs a refinement algorithm based on maximum-flow computations between pairs of blocks.
Since flow algorithms find minimum $s$-$t$ cuts, this refinement technique does not suffer the drawbacks of move-based approaches.

The flow-based refinement framework is a generalization of the approach used in the graph partitioner KaFFPa~\cite{ss-emgpa-11}.
To refine a pair of blocks of a $k$-way partition, KaHyPar first extracts a subhypergraph induced by a set of vertices around the cut of these blocks.
This subhypergraph is then transformed into a graph-based flow network, using techniques due to Lawler~\cite{l-cph-73}, and Liu and Wong~\cite{hw-nfbmp-98}, on which a maximum flow is computed.
The vertices of the subhypergraph are reassigned according to the corresponding minimum cut.
The size of the subhypergraph (and thus the size of the flow network) is chosen adaptively, depending on the outcome of the previous refinement. 
While larger flow networks may produce better but potentially imbalanced solutions, the smallest flow network guarantees a balanced partition.
We further discuss KaHyPar and its flow-based refinement framework in Section~\ref{sec:prelim}.

HyperFlowCutter (HFC)~\cite{ghw-efbhba-esa19} computes bipartitions of unweighted hypergraphs by solving a sequence of incremental maximum flow problems.
Its advantage over computing a single maximum flow is that it does not reject almost balanced solutions, but systematically trades cut-size for increased balance.
ReBaHFC~\cite{ghw-efbhba-esa19} uses HFC as postprocessing to improve an initial bipartition computed with PaToH.
HFC computes unit-capacity flows directly on the hypergraph (i.e., without using a graph-based flow network) using a technique of Pistorius and Minoux~\cite{pm-aidlm-03} extended to Dinic's flow algorithm~\cite{d-aspmf-70}.

\subparagraph*{Contribution and Outline}

Multilevel refinement algorithms must be able to handle both weighted hyperedges and weighted vertices because vertices and hyperedges are aggregated during the coarsening phase.
In this work, we improve KaHyPar's flow-based refinement framework -- with regard to both running time and solution quality -- by using a \emph{weighted} version of HFC instead of the maximum flow computations on differently-sized graph-based flow networks.
After introducing notation and briefly presenting additional details about KaHyPar and HFC refinement in Section~\ref{sec:prelim}, we discuss how HFC can simulate an approach of KaHyPar and KaFFPa for balancing partitions, extend HFC's existing balancing approach to weighted instances, and propose a heuristic for guiding its incremental maximum flow problems in Section~\ref{sec:whfc}.
In Section~\ref{sec:flows_on_hypergraphs}, we present our approach for computing maximum flows on weighted hypergraphs, generalizing the technique of Pistorius and Minoux~\cite{pm-aidlm-03} to weighted hypergraphs and arbitrary flow algorithms.

In our experiments (Section~\ref{sec:experiments}), we compare two configurations of our new approach with KaHyPar-MF,
and seven other partitioning algorithms on a large benchmark set containing hypergraphs from the VLSI, SAT solving, and
scientific computing community~\cite{hs-icshp-17}.
While our first configuration, KaHyPar-HFC, computes slightly better solutions than KaHyPar-MF using significantly less running
time, the second configuration, KaHyPar-HFC*, computes solutions of significantly better quality and is still slightly faster than KaHyPar-MF.
Furthermore, in terms of solution quality, both configurations outperform all other competing partitioners.
We conclude the paper in Section~\ref{sec:conclusion} and suggest future work.

\section{Preliminaries}\label{sec:prelim}

\subparagraph*{Hypergraphs.}
A \emph{hypergraph} $H=(V,E)$ consists of a set of vertices $V$ and a set of hyperedges $E$, where a hyperedge $e$ is a subset of the vertices $V$.
Additionally, we associate weights $\omega \colon E \to \mathbb{N}^+$, $\varphi \colon V \to \mathbb{N}^+$ with the hyperedges and vertices.
A vertex $v \in V$ is \emph{incident} to hyperedge $e \in E$ if $v \in e$.
The vertices incident to a hyperedge $e$ are called the \emph{pins} of $e$.
We denote the pin $u$ in hyperedge $e$ as $(u,e)$. 
By $H[V'] = (V', \{ e \cap V' \mid e \in E \})$, we denote the hypergraph induced by the vertex set $V'$.
The \emph{star expansion} of $H$ represents the hypergraph as a bipartite graph $G=(V \dot\cup E, \{ (v,e) \in V \times E \mid v \in e\})$ with bipartite node set $V \dot\cup E$ and an edge for every pin. 
To avoid confusion, we use the terms vertices, hyperedges, and pins for hypergraphs, and the terms nodes and edges for graphs.
We extend functions to sets using $f(X) = \sum_{x \in X} f(x)$ for some function $f$.

\subparagraph*{Hypergraph Partitioning.}
A $k$-way partition $\pi(H)$ of a hypergraph $H=(V,E)$ is a partition of its vertices into non-empty disjoint \emph{blocks} $V_1, \dots, V_k \subset V$, \ie, $\bigcup_{i=1}^k V_i = V$, $V_i \neq \emptyset$ for $i = 1,\dots,k$ and $V_i \cap V_j = \emptyset$ for $i \neq j$.
For some parameter $\epsilon \in [0, 1)$ we call $\pi(H)$ $\epsilon$-balanced, if each block $V_i$ satisfies the \emph{balance constraint} $\varphi(V_i) \leq (1+\epsilon) \frac{\varphi(V)}{k}$.
Let $\Lambda(e) = \{ V_i \in \pi(H) \mid V_i \cap e \neq \emptyset \}$ denote the blocks that are connected by hyperedge $e \in E$. 
The \emph{connectivity} of a hyperedge $e$ is defined as $\lambda(e) := |\Lambda(e)|$.
Given parameters $\epsilon$ and $k$, and an input hypergraph $H$, the balanced $k$-way hypergraph partitioning problem asks for an $\epsilon$-balanced $k$-way partition of $H$ that minimizes the \emph{connectivity-metric} $\sum_{e \in E} \omega(e) (\lambda(e) - 1)$.

\subparagraph*{Maximum Flows.}
A flow network is a symmetric, directed graph $\mathcal{N} = (\mathcal{V},\mathcal{E})$ with two disjoint non-empty \emph{terminal} node sets $S,T\subsetneq \mathcal{V}$, the source and target node set, as well as a capacity function $c \colon \mathcal{E} \to \mathbb{N}_0$.
A flow is a function $f \colon \mathcal{E} \to \mathbb{Z}$ subject to the \emph{capacity constraint} $f(e) \leq c(e)$ for all edges $e$, \emph{flow conservation} ${\sum_{(u,v)\in \mathcal{E}} f(u,v) = 0}$ for all non-terminal nodes $v$, and \emph{skew symmetry} $f(u,v)=-f(v,u)$ for all edges~$(u,v)$. 
The \emph{value} of a flow ${|f| := \sum_{s \in S, (s,u)\in \mathcal{E}} f(s,u)}$ is the amount of flow leaving $S$.
The \emph{residual capacity} $r_f(e) := c(e) - f(e)$ is the additional amount of flow that can pass through $e$ without violating the capacity constraint.
The residual network with respect to $f$ is the directed graph $\mathcal{N}_f = (\mathcal{V},\mathcal{E}_f)$, where $\mathcal{E}_f := \{e \in \mathcal{E} | r_f(e) > 0\}$.
A node $v$ is \emph{source-reachable} if there is a path from $S$ to $v$ in $\mathcal{N}_f$, it is \emph{target-reachable} if there is a path from $v$ to $T$ in $\mathcal{N}_f$.
We denote the source-reachable and target-reachable nodes by $S_r$ and $T_r$, respectively.
An \emph{augmenting path} is an $S$-$T$ path in $\mathcal{N}_f$.
The flow $f$ is a \emph{maximum flow} if $|f|$ is maximal of all possible flows in $\mathcal{N}$. 
This is the case if and only if there is no augmenting path in $\mathcal{N}_f$.
An $S$-$T$ edge cut is a set of edges whose removal disconnects $S$ and $T$.
The value of a maximum flow equals the weight of a minimum-weight $S$-$T$ edge cut~\cite{ff-mftn-56}.
The \emph{source-side cut} consists of the edges from $S_r$ to $\mathcal{V} \setminus S_r$ and the \emph{target-side cut} consists of the edges from $T_r$ to $\mathcal{V} \setminus T_r$.
The bipartition $(S_r, \mathcal{V} \setminus S_r)$ is induced by the source-side cut and $(\mathcal{V} \setminus T_r, T_r)$ is induced by the target-side cut.
Note that $\mathcal{V} \setminus S_r \setminus T_r$ is not necessarily empty.
We also call $S_r$ and $T_r$ the \emph{cutsides} of a maximum flow.

\subparagraph*{Maximum Flows on Hypergraphs.}
Lawler~\cite{l-cph-73} uses maximum flows to compute minimum $S$-$T$ hyperedge cuts.
On the star expansion, the construction to model node capacities as edge capacities~\cite{amo-nf-93} is applied to the hyperedge-nodes.
A hyperedge $e$ is expanded into an \emph{in-node} $e_i$ and an \emph{out-node} $e_o$ joined by a directed \emph{bridge edge} $(e_i, e_o)$ with capacity $c(e_i, e_o) = \omega(e)$.
For every pin $u \in e$ there are two directed \emph{external edges} $(u, e_i), (e_o, u)$ with infinite capacity.
The transformed graph is called the \emph{Lawler network}.
A minimum $S$-$T$ edge cut in the Lawler network consists only of bridge edges, which directly correspond to $S$-$T$ cut hyperedges in $H$.
Via the Lawler network, the above notions translate naturally from graphs to hypergraphs, and we use the same terminology and notation for hypergraphs.

\subparagraph*{The KaHyPar Framework.}
Since our algorithm is integrated into the KaHyPar framework, we briefly review its core components and outline its $k$-way flow-based refinement.
In contrast to traditional multilevel HGP algorithms that contract matchings or clusterings and therefore work with a coarsening hierarchy of $O(\log n)$ levels, KaHyPar removes only a single vertex between two levels, resulting in almost $n$ levels.
Coarsening is restricted to clusters found with a community detection algorithm~\cite{hs-icshp-17}.
Initial partitions of the coarsest hypergraph are computed using a portfolio of simple algorithms~\cite{shhmss-hplrb-16}.
During uncoarsening, it employs a combination of localized FM local search~\cite{ahss-ehpa-17,shhmss-hplrb-16} and flow-based refinement~\cite{kahypar-mf-19}.

Given a $k$-way partition $\pi(k) = \{V_1,...,V_k\}$ of a hypergraph $H = (V, E, \varphi, \omega)$, the flow-based refinement works on pairs $(V_i,V_j)$ of blocks that share cut hyperedges.
The blocks are scheduled for refinement as long as this improves the solution (better connectivity, or equal connectivity and less imbalance).
To construct a flow problem for a pair of blocks, the algorithm performs two randomized weight-constrained breadth-first searches (BFS) restricted to $H[V_i]$ and $H[V_j]$, respectively.
The BFSs are initialized with the vertices of $V_i$ (resp. $V_j$) that are incident to hyperedges in the cut between $V_i$ and $V_j$.
The first BFS stops if the weight of the visited vertices would exceed $(1+ \alpha \cdot \epsilon) \lceil \frac{\varphi(V)}{k} \rceil - \varphi(V_j)$, where the scaling parameter $\alpha$ is used to control size of the flow problem.
The second BFS visits the vertices of $V_j$ analogously.
The hypergraph induced by the visited vertices is then used to build the Lawler network, with size optimizations due to Liu and Wong~\cite{hw-nfbmp-98} and Heuer~\cite{kahypar-mf-19}.
A minimum $S$-$T$ cut in the Lawler network induces a bipartition of $H[V_i \cup V_j]$.
If the flow computation resulted in an $\epsilon$-balanced partition, the improved solution is accepted and $\alpha$ is increased to $\min(2 \alpha, \alpha')$ for a predefined upper bound of $\alpha' = 16$. 
Otherwise, $\alpha$ is decreased to $\max( \alpha/2 , 1)$. 
This scaling scheme continues until a maximal number of rounds is reached or a feasible partition that did not improve the cut is found.
KaHyPar runs flow-based refinement on exponentially spaced levels, \ie, after $2^i$ uncontractions for increasing $i$, since flow-based refinement is too expensive to be run after every uncontraction.

\subparagraph*{ReBaHFC.}
ReBaHFC avoids the need for Lawler networks and the corresponding size optimizations~\cite{hw-nfbmp-98,kahypar-mf-19} by directly constructing a \emph{flow hypergraph} by contracting vertices not visited by a BFS to $S$ or $T$.
Furthermore, it does not require adaptive rescaling because HFC guarantees balanced partitions.

\section{Weighted HyperFlowCutter}\label{sec:whfc}
To keep this paper self-contained, we briefly explain the core HFC algorithm in Section~\ref{sec:core_algorithm}.
Subsequently, we discuss further details of multilevel refinement with HFC in Section~\ref{sec:hfc_refinement} and our approach for adapting two balancing strategies in Section~\ref{sec:balancing}.

\subsection{The Core Algorithm}\label{sec:core_algorithm}

\begin{figure}[tb]
	\begin{subfigure}[t]{.45\linewidth}
		\includegraphics[width=\linewidth, page=1]{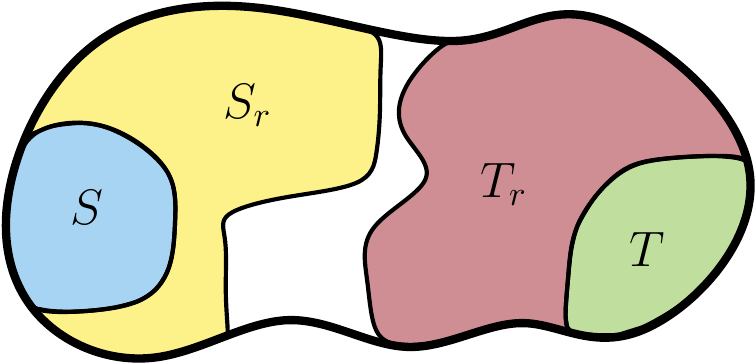}
		\caption{Compute minimum $S$-$T$ cuts.}\label{fig:fc_initialstate}
	\end{subfigure}
	\hfill
	\begin{subfigure}[t]{.45\linewidth}
		\includegraphics[width=\linewidth, page=2]{illustrations/flowcutter.pdf}
		\caption{Add $S_r$ to $S$ and choose a piercing vertex.}\label{fig:fc_pierce}
	\end{subfigure}
	\caption{
		Flow augmentation and computing $S_r, T_r$ in Fig.\ref{fig:fc_initialstate}; adding $S_r$ to $S$ and piercing the source-side cut in Fig.\ref{fig:fc_pierce}. $S$ in blue, $S_r \setminus S$ in yellow, $T$ in green, $T_r\setminus T$ in red, $V \setminus S_r, \setminus T_r$ in white.
		Taken from Gottesbüren \etal~\cite{ghw-efbhba-esa19} with minor adaptations.
	}\label{fig:illuHFC}
\end{figure}

The core HFC algorithm computes bipartitions with monotonically increasing cut size and balance by solving a sequence of incremental maximum flow problems, until both blocks satisfy the balance constraint.
Given some initial terminal vertex sets $S$ and $T$, HFC first computes a maximum flow and derives the cutsides $S_r$ and $T_r$.
If either the source-side cut or the target-side cut induces blocks that fulfill the balance constraint, the algorithm terminates.
Otherwise, it adds all vertices on the smaller cutside to its corresponding terminal side, \ie, $S := S_r$ if $|S_r| \leq |T_r|$ and $T := T_r$ otherwise.
Then, one additional vertex -- the \emph{piercing vertex} -- is added to the terminal side and the previous flow is augmented to respect the new terminals.
This ensures that HFC finds a different cut with each new maximum flow.
By growing the smaller side, the algorithm ensures that it finds a balanced partition after at most $|V|$ iterations.
Figure~\ref{fig:illuHFC} illustrates the HFC phases.
HFC selects piercing vertices for $S$ from the boundary vertices of the source-side cut, \ie, the vertices incident to hyperedges in the source-side cut that are not already contained in $S$.
Analogously, the candidates for $T$ are the boundary vertices of the target-side cut.
Note that the previous flow still satisfies the flow constraints, and only the piercing vertex can break the maximality of the flow.
HFC prefers piercing vertices that maintain the maximality of the current flow, \ie, do not create an augmenting path.
This strategy is called the \emph{avoid-augmenting-paths} piercing heuristic.
Adding a vertex $u$ to $S$ creates an augmenting path if and only if $u \in T_r$.

\subsection{Multilevel Refinement Using HyperFlowCutter}\label{sec:hfc_refinement}

For multilevel $k$-way refinement with HyperFlowCutter we use the block-pair scheduling of KaHyPar-MF and the flow model construction of KaHyPar-MF and ReBaHFC.
Unlike ReBaHFC, we do not explicitly contract unvisited vertices to $S$ (resp. $T$).
Instead, we build the flow hypergraph during the BFS and mark pins that will not be in the flow hypergraph as terminals.
Hyperedges containing both terminals are removed as they cannot be eliminated from the cut.
Subsequently, our weighted HFC algorithm is run on the weighted flow hypergraph.
We stop once the flow exceeds the weight of the remaining hyperedges from the original cut.
The difference between the weight of the original cut hyperedges and the flow value equals the decrease in connectivity.

\subparagraph*{Distance-Based Piercing.}
We can use the original cut to guide HFC.
To avoid that bad piercing decisions make it impossible for HFC to recover parts of the original cut, we use BFS-distances from the original cut as an additional piercing heuristic.
We prefer larger distances, secondary to avoiding augmenting paths.
Vertices from the other side of the original cut are rated with distance $-1$, \ie, chosen only after one side has been entirely added to the corresponding terminal vertices.
This is similar to the flow network rescaling of KaHyPar, as we first use vertices as terminals that could only be contained in larger flow networks.
We maintain the boundary vertices in a bucket priority queue and select candidates uniformly at random from the highest-rated non-empty bucket.
New terminal vertices are removed lazily.

\subsection{Improved Balance}\label{sec:balancing}

KaHyPar uses both flows and FM local search to refine a partition.
Because FM only considers moves that maintain the balance constraint, partitions with small imbalance tend to give FM more leeway for improving the current solution.
In this section, we discuss two approaches to improve the balance during HFC refinement.
These can also improve the solution quality, since HFC would otherwise trade better balance for a larger cut.

\subparagraph*{Keep Piercing.}

Given a maximum flow and minimum cut, finding a most balanced cut of the same weight is NP-hard~\cite{b-m-10}.
All vertices of a strongly connected component (SCC) of the residual network belong to the same side in a minimum cut.
Hence, finding a most balanced minimum cut corresponds to a knapsack problem with partial order constraints induced by the directed acyclic graph (DAG) obtained from contracting all SCCs~\cite{pq-osamc-82}.
Each topological ordering of the DAG corresponds to a series of minimum cuts. 
KaHyPar computes several random topological orderings to find a cut with the same weight and less imbalance, which considerably improved solution quality~\cite{kahypar-mf-19}.

Using the avoid-augmenting-paths piercing heuristic, we can perform such a sweep without the need to explicitly construct the DAG and to compute a topological ordering.
Instead of stopping at the first balanced partition, we continue to pierce as long as no augmenting path is created.
Since this process is fast and piercing decisions are randomized, we repeat it several times and select the partition with the smallest imbalance.

\subparagraph*{Reassigning Isolated Vertices.}
A vertex $v \notin S \cup T$ is called \emph{isolated} if all of its incident hyperedges have pins in both $S$ and $T$.
An isolated vertex $v$ can be moved without affecting the cut, because $v \notin S_r \cup T_r$ and its incident hyperedges remain in both the source- and target-side cut over the course of the algorithm.
Using the terminology above, $v$ is not affected by partial order constraints.
This reduces the optimal assignment problem for isolated vertices to a subset sum problem.
With unweighted vertices, we can easily distribute the total weight $|L|$ of a set $L$ of isolated vertices among the two sides to achieve optimum balance.
Introducing vertex weights makes the subset sum problem non-trivial, because arbitrary divisions of the total weight $\varphi(L)$ are not necessarily possible.
Since isolated vertices remain isolated, the problem instances are incremental.
To solve the problem, we use the pseudo-polynomial dynamic program (DP) for subset sum~\cite[Section 35.5]{clrs-ia-01}.
It maintains a lookup table of partition weights that are summable with isolated vertices.
After obtaining a new cut, we update the DP table to incorporate potential new isolated vertices.
For each new isolated vertex $v$, we iterate over the subset sums $x$ in the table and insert $\varphi(v) + x$ if it was not yet a subset sum.
As an optimization, we maintain a list of ranges that are summable (consecutive entries).
The balance check takes constant time per range.
To merge ranges efficiently, we store a pointer from each DP table entry to its range in the list.
When a new subset sum $x$ is obtained, we check whether $x-1$ and $x+1$ are also subset sums, and extend or merge ranges as appropriate.
Since entries in the DP table cannot be reverted easily, we do not add isolated vertices to the DP after the first balanced partition is found.

As the DP is only pseudo-polynomial, its running time may become prohibitive on instances with very large vertex weights.
In our experiments, non-unit vertex weights are only the result of contractions during coarsening.
Hence, the DP is polynomial in the size of the unit-weight input hypergraph.
We plan to implement a simple classifier to decide whether the running time can be harmful, and deactivate the DP accordingly.

\section{Maximum Flows on Weighted Hypergraphs}\label{sec:flows_on_hypergraphs}

\newcommand{\wf}{\widetilde{f}}

\begin{figure}[tb]
	\includegraphics[width=\linewidth]{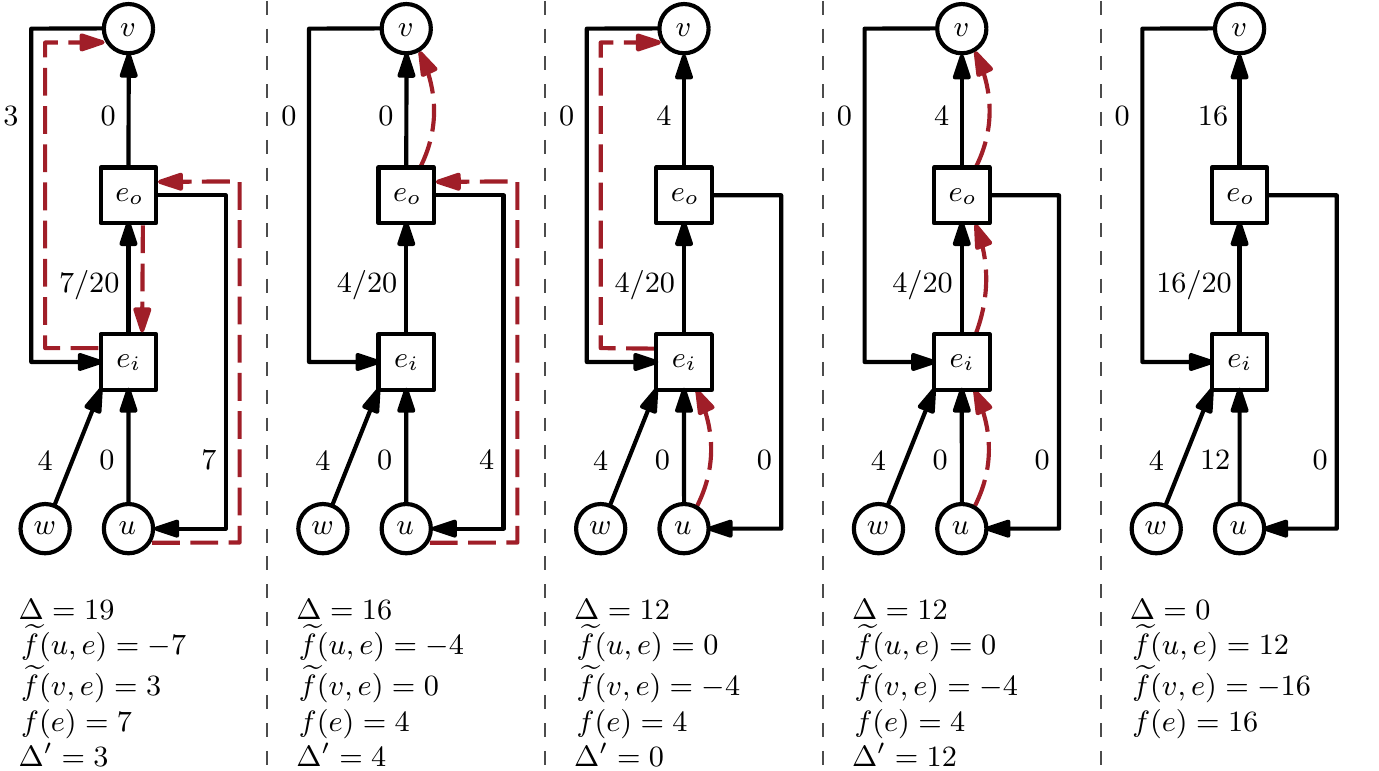}
	\caption{
		Example illustrating the four steps for pushing $\Delta = 19$ units of flow from $u$ via hyperedge $e = \{u,v,w\}$ to $v$.
		Black edges show the direction of the flow, dashed red arrows the direction we want to push flow in the Lawler network.
		The edge $(e_0, w)$ is omitted for readability.
		Current values of $\Delta$, $\wf(u,e)$, $\wf(v,e)$, $f(e)$, and $\Delta'$ are shown at the bottom for each state.
	}
	\label{fig:route_flow}
\end{figure}

In this section, we present our technique for computing maximum flows on weighted hypergraphs.
It generalizes the approach of Pistorious and Minoux~\cite{pm-aidlm-03} (which computes maximum flows directly on unweighted hypergraphs using the Edmonds-Karp algorithm~\cite{ek-tiaenf-72}) to weighted hypergraphs and to arbitrary flow algorithms.

Let $P$ denote the set of pins, and let $\wf(u,e) \colon P \to \mathbb{Z}$ denote the amount of flow that vertex $u$ sends into hyperedge $e$.
Negative values indicate that $u$ receives flow from $e$.
Let $\wf(u, e)^+ := \max(\wf(u, e), 0)$ denote the net flow $u$ sends into $e$ and $\wf(u, e)^- := \max(-\wf(u, e), 0)$ the net flow $u$ receives from $e$.
Then, $f(e)$ can also be written as $\sum_{u \in e} \wf(u,e)^+ = \sum_{u \in e} \wf(u,e)^-$.
We can push up to $c(e) - f(e) + \wf(u,e)^- + \wf(v,e)^+$ flow from $u$ via $e$ to another pin $v \in e$.
The advantage of implementing flow algorithms directly on the hypergraph is the ability to identify and skip cases in which we cannot push any flow.
If $c(e) - f(e) = 0$ and $\wf(u,e) \geq 0$, we only need to iterate over the pins $v\in e$ with $\wf(v,e) > 0$.
A graph-based flow algorithm on the residual Lawler network would scan the edge $(e_i, v)$ for \emph{every} pin $v \in e$.
For unweighted hypergraphs, each $e \in E$ has at most one pin $v \in e$ with $\wf(v,e) > 0$, which can be stored in a separate array of size $|E|$.
Since HFC needs to compute flows in forward and backward direction, an additional array is needed for pins $v' \in e$ with $\wf(v',e) < 0$.
A weighted hyperedge can have many such pins, which would require arrays of size $|P|$.
Instead, we divide the pins of $e$ into subranges $\wf(v,e) \{ < 0, = 0, > 0\}$.
When the sign of $\wf(v,e)$ changes, we insert pin $v$ into the correct subrange by performing swaps with the range boundaries.

In addition to scanning hyperedges like vertices, pushing flow over a hyperedge is the elementary operation necessary to implement any flow algorithm on hypergraphs.
Let $\Delta$ be the amount of flow to push.
We update the values $\wf(u,e), \wf(v,e),$ and $f(e)$ in four steps (see Figure~\ref{fig:route_flow} for an example).
These steps correspond to paths $p_1 = (u, e_o, e_i, v)$, $p_2 = (u, e_o, v)$, $p_3 = (u, e_i, v)$, $p_4 = (u, e_i, e_o, v)$ in the residual Lawler network.
The order of these steps is important to correctly update $f(e)$.
First, we push $\Delta' := \min(\Delta, \wf(u,e)^-, \wf(v,e)^+)$ along $p_1$ by setting $f(e) := f(e) - \Delta'$, $\wf(u,e) := \wf(u,e) + \Delta'$, $\wf(v,e) := \wf(v,e) - \Delta'$, and $\Delta := \Delta - \Delta'$.
Then, we push $\Delta' := \min(\Delta, \wf(u,e)^-)$ along $p_2$, by updating $\wf(u,e), \wf(v,e),$ and $\Delta$ as before.
Note that we do not update $f(e)$ since the bridge edge $(e_i, e_o)$ is not in $p_2$.
Analogously to $p_2$, we push $\Delta' := \min(\Delta, \wf(v,e)^+)$ along $p_3$.
Finally, we push the remaining $\Delta$ along $p_4$ and update $f(e), \wf(u,e), \wf(v,e),$ and $\Delta$ as for $p_1$.

Note that the Lawler network is just used as a means of illustration.
In our implementation, we only update the $\wf(u,e), \wf(v,e)$, and $f(e)$ values as shown at the bottom of Figure~\ref{fig:route_flow}.

\subparagraph*{Flow Algorithm.}
We chose to implement Dinic's flow algorithm~\cite{d-aspmf-70} with capacity scaling, because the flow problems for HFC refinement on our benchmark set predominantly have a small diameter (due to the BFS-based construction).
Dinic's algorithm consists of two alternating phases: assigning hop-distance labels to vertices by performing a BFS on the residual network, and using DFS to find edge-disjoint augmenting paths with distance labels increasing by one along the path.
In addition to vertex distance labels, we maintain hyperedge distance labels.
For a vertex $u$, we only traverse those incident hyperedges $e$ whose distance label matches that of $u$.
The pins $v \in e$ are only traversed if the distance of $v$ is $1$ plus the distance of $e$.
To distinguish the case that we can only push flow to pins $v$ with $\wf(v,e) > 0$, we actually maintain two different distance labels per hyperedge.

\subparagraph*{Optimizations.}
It suffices to initialize the BFS and DFS with the last piercing vertex $p$ of a side, since only $p$ can lead to newly reachable vertices.
When Dinic's algorithm terminates, we already know $S_r$ or $T_r$ (depending on which side was pierced) and thus only compute the reachable vertices of the other side.
While computing this set, we also compute the corresponding distance labels, so that the next flow computation can directly start with the DFS phase.
Additionally, we infer the sets $S_r, T_r, S,$ and $T$ from the distance labels.

\section{Experimental Evaluation}\label{sec:experiments}

The C++17 source code for Weighted HyperFlowCutter\footnote{\url{https://github.com/larsgottesbueren/WHFC}} and KaHyPar-HFC\footnote{\url{https://github.com/SebastianSchlag/kahypar}} are available as open-source software.
Experiments are performed sequentially on a cluster of Intel Xeon E5-2670 Sandy Bridge nodes with two Octa-Core processors clocked at 2.6 GHz with 64 GB RAM, 20~MB L3- and 8$\times$256 KB L2-Cache, using only one core of a node.

We consider two configurations which differ in the constraint for vertices from $V_i$ in the flow hypergraph.
KaHyPar-HFC uses $\frac{1}{5} \cdot \varphi(V_i)$ (as used for RebaHFC~\cite{ghw-efbhba-esa19}), while KaHyPar-HFC* uses $(1+ 16 \cdot \epsilon) \lceil \frac{\varphi(V)}{k} \rceil - \varphi(V_j)$ (as used for KaHyPar-MF~\cite{kahypar-mf-19}).
In Appendix~\ref{sec:component_eval}, we assess the impact of the algorithmic components on the solution quality of KaHyPar-HFC.
Both configurations use all components.

\subparagraph*{Benchmark Set.}
We use a comprehensive benchmark set of real-world hypergraphs compiled by Schlag~\cite{hs-icshp-17}.\footnote{\url{https://algo2.iti.kit.edu/schlag/sea2017/}}
It consists of 488 unit-weight hypergraphs from four sources: the ISPD98 VLSI Circuit Benchmark Suite~\cite{a-tispd-98} (ISPD98, 18 hypergraphs), the DAC 2012 Routability-Driven Placement Benchmark Suite~\cite{naslw-tdacr-12} (DAC, 10), the SuiteSparse Matrix Collection~\cite{dh-tufsm-11} (SPM, 184) and the international SAT Competition 2014~\cite{sat2014} (Literal, Primal, Dual, 92 hypergraphs each).
Refer to Table~\ref{table:instance_stats} in Appendix~\ref{appendix:instances} for hypergraph sizes and refer to~\cite{hs-icshp-17} for information on how the hypergraphs were derived.
We compute partitions for $\epsilon = 3\%$ and $k \in \{2,4,8,16,32,64,128\}$.
Each combination of a hypergraph and a value of $k$ constitutes one \emph{instance}, resulting in a total of $3416$ instances.

\subparagraph*{Methodology.}
In addition to the KaHyPar configurations, we consider the state-of-the-art partitioners hMetis~\cite{kaks-mhpav-99,kk-mkwhp-99} in both recursive bisection (-R) and direct k-way (-K) mode, PaToH~\cite{patoh} with default (-D) and quality preset (-Q), Zoltan-AlgD~\cite{scs-rbcmh-19}, Mondriaan~\cite{vb-atddd-05}, as well as HYPE~\cite{mmber-hypem-18}.
We excluded MLPart~\cite{ahk-mcp-98} and ReBaHFC~\cite{ghw-efbhba-esa19} because they are restricted to bipartitioning, and Par$k$way~\cite{tk-pmahp-08} because we were not able work with the current version~\footnote{\url{https://github.com/parkway-partitioner/parkway}}.
The code hangs on many instances, e.g. it did not finish within 8 hours on a small instance that took the other partitioners around 20 seconds.

For each instance and partitioner, we perform ten runs with different random seeds.
The only exception is HYPE~\cite{mmber-hypem-18} which produced worse solutions when randomized~\cite{kahypar-mf-19}.
Hence, we report only one non-randomized run of HYPE.
Running times and connectivity values per instance are aggregated using the arithmetic mean, while running times across instances are aggregated using the geometric mean to give instances of different sizes a comparable influence.
To compare running times we use a combination of a scatter plot, which shows the arithmetic mean time per instance, and a box-and-whiskers plot~\cite{tukey1977box}.
Because small running times are more frequent, we use a fifth-root scale~\cite{cuberoots} on the y-axis.
Runs that exceeded the 8 hour time limit count as 8 hours in the reported aggregates and plots.

For comparing solution quality we use performance profiles~\cite{performanceprofiles}.
Let $\mathcal{A}$ denote a set of algorithms, $\mathcal{I}$ a set of instances and $\obj(a,i)$ denote the objective value computed by $a \in \mathcal{A}$ on $i \in \mathcal{I}$ -- in our case the arithmetic mean connectivity of 10 runs.
The performance ratio
\[ r(a,i) = \frac{\obj(a,i)}{ \min \{\obj(a',i) \, \mid \, a' \in \mathcal{A} \} } \]
indicates by what factor $a$ deviates from the best solution on instance $i$.
In particular, algorithm $a$ found the best solution on instance $i$ if $r(a,i) = 1$.
The performance profile
\[ 
\rho_a \colon [1,\infty) \to [0,1], \tau \mapsto \frac{| \{i \in \mathcal{I} \mid r(a,i) \leq \tau \} | } {|\mathcal{I}|}
\]
of $a$ is the fraction of instances for which it is within a factor of $\tau$ from the best solution.
Runs that did not finish within the time limit or resulted in an error (balance violation or crash) are excluded.
If this concerns all runs of an algorithm on an instance, we report the corresponding fractions as the steps at special symbols (\ding{99}, \ding{56} respectively).


\subsection{Comparison with KaHyPar-MF}

\begin{figure}[tb]
	\begin{subfigure}[t]{.5\linewidth}
	\includegraphics[width=\linewidth]{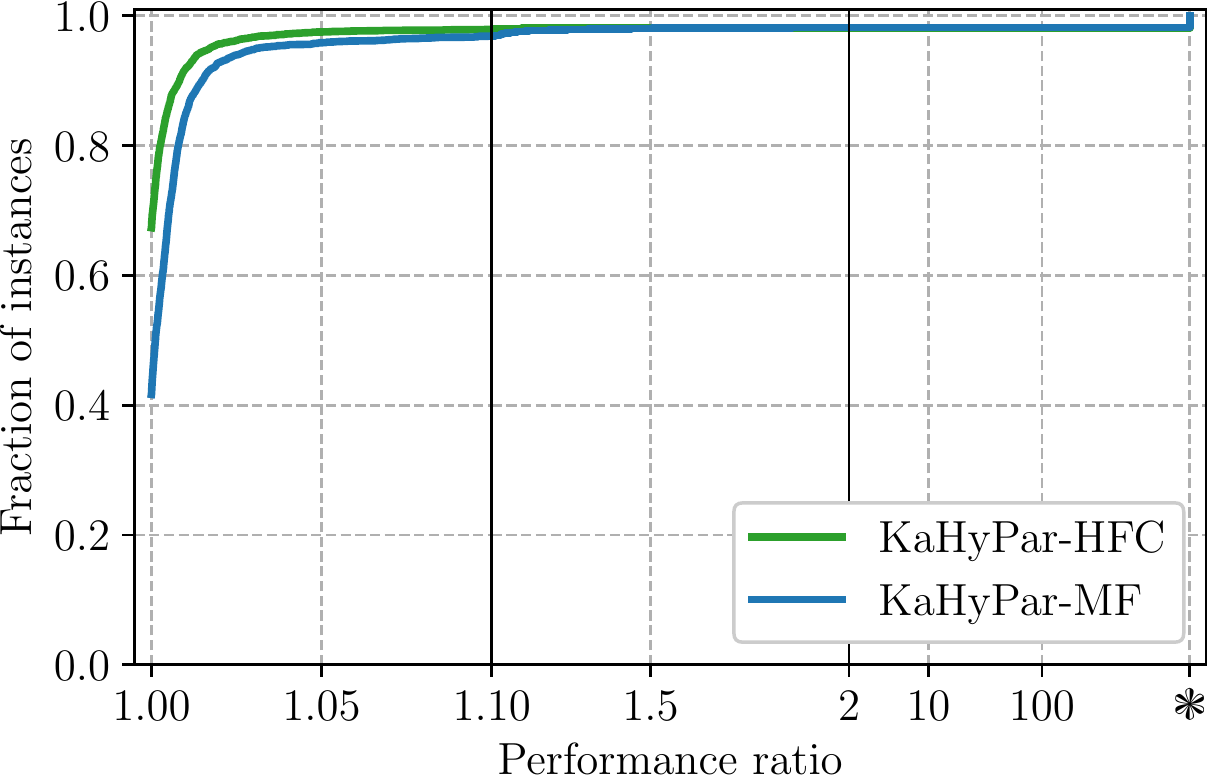}
	\end{subfigure}
	\hfill
	\begin{subfigure}[t]{.5\linewidth}
		\includegraphics[width=\linewidth]{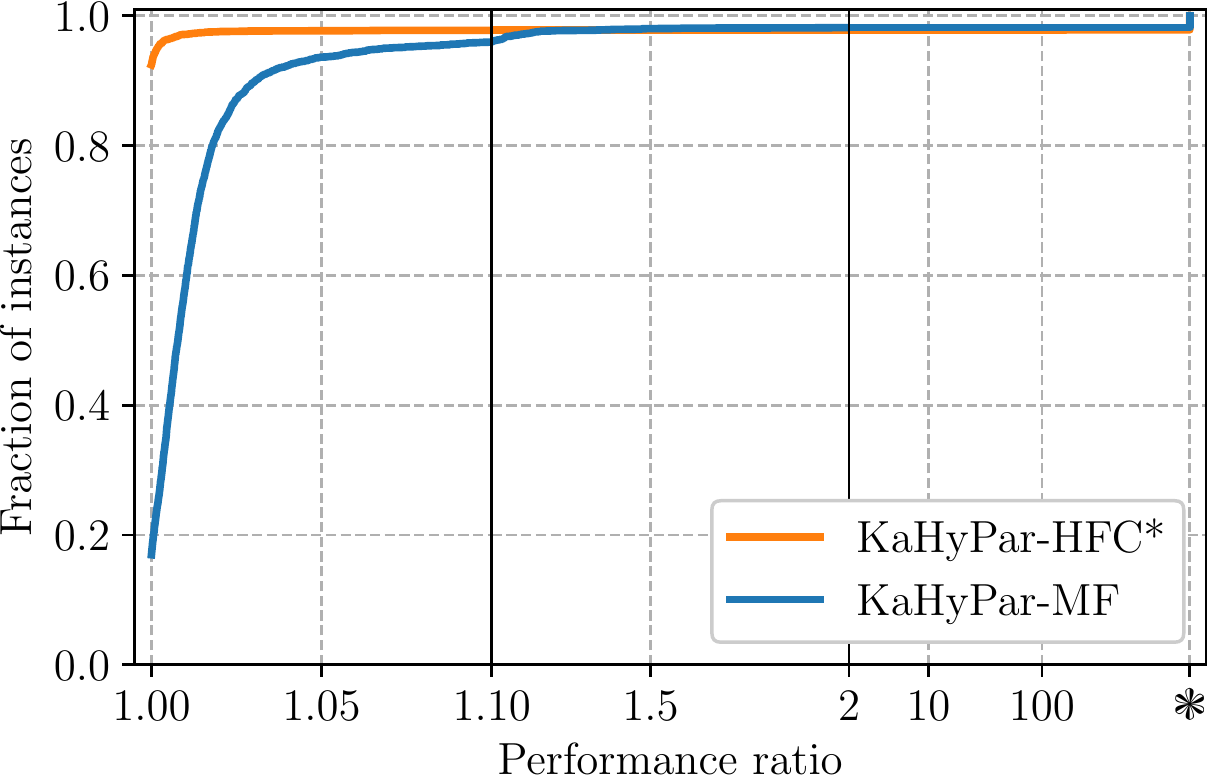}
	\end{subfigure}
	\caption{Performance profiles comparing our new variants with KaHyPar-MF.}\label{fig:kahypar_comparisons}
\end{figure}

\begin{figure}[tb]
	\begin{center}
		\includegraphics[width=.7\linewidth]{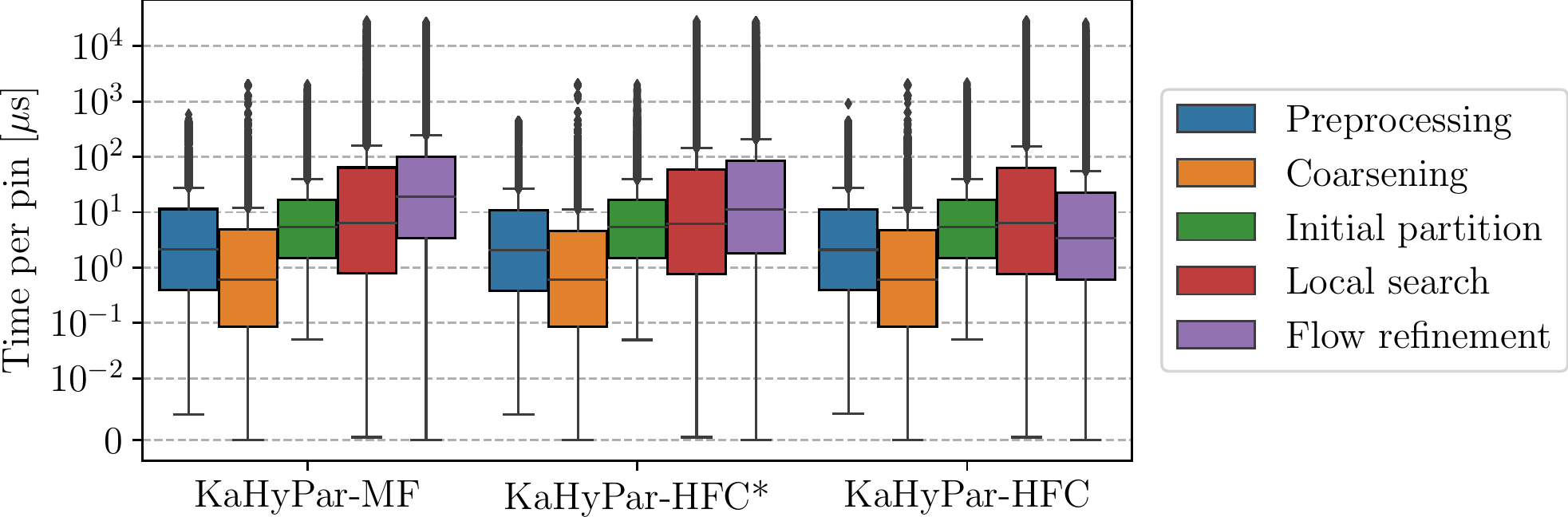}
		\caption{Box plots comparing running times (in $\mu$s per pin) of the different phases of KaHyPar.}\label{fig:kahypar_phases}
	\end{center}
\end{figure}

KaHyPar-HFC computes solutions with better, equal, or worse quality than KaHyPar-MF on 1933, 367, 1056 instances, respectively.
On the remaining 60 instances neither finished within the time limit.
As Figure~\ref{fig:kahypar_comparisons} (left) shows, the performance ratios are consistently better, though not by a large margin.
The median fraction of flow-based refinement time of KaHyPar-HFC vs KaHyPar-MF is 0.18, the 75th percentile is 0.26, and the 90th percentile is 0.51.
Hence, the flow-based refinement of KaHyPar-HFC is significantly faster than that of KaHyPar-MF.
Flow-based refinement constitutes about 40\% of KaHyPar-MF's overall running time~\cite{kahypar-mf-19}.
With a mean overall running time of 44.84s, KaHyPar-HFC is about 33\% faster than KaHyPar-MF at 67.07s.
The improved running time is partially due to faster flow computation and partially due to smaller flow hypergraphs.
Since KaHyPar-HFC uses smaller flow hypergraphs, the improved solution quality can be attributed to the HyperFlowCutter approach.
With 62.49s, KaHyPar-HFC* is moderately faster than KaHyPar-MF and computes solutions of better, equal, or worse quality on 2776, 381, 198 instances, respectively (with 61 instances on which neither the -HFC* nor the -MF variant finished within the time limit).
Hence, the faster flow computation more than compensates the additional work incurred by HFC.
Figure~\ref{fig:kahypar_phases} shows box plots for the different phases of KaHyPar.
The running times of preprocessing, coarsening, and initial partitioning remain unchanged, as they are not influenced by the refinement phase.
During the refinement phase, local search and flow-based improvement both modify the solution and thus influence one another.
The plots show that the running time of local search remains largely unchanged, while our variants reduce the running time of flow-based refinement.

\subsection{Comparison with other Partitioners}

\begin{figure}[tb]
	\includegraphics[width=\linewidth]{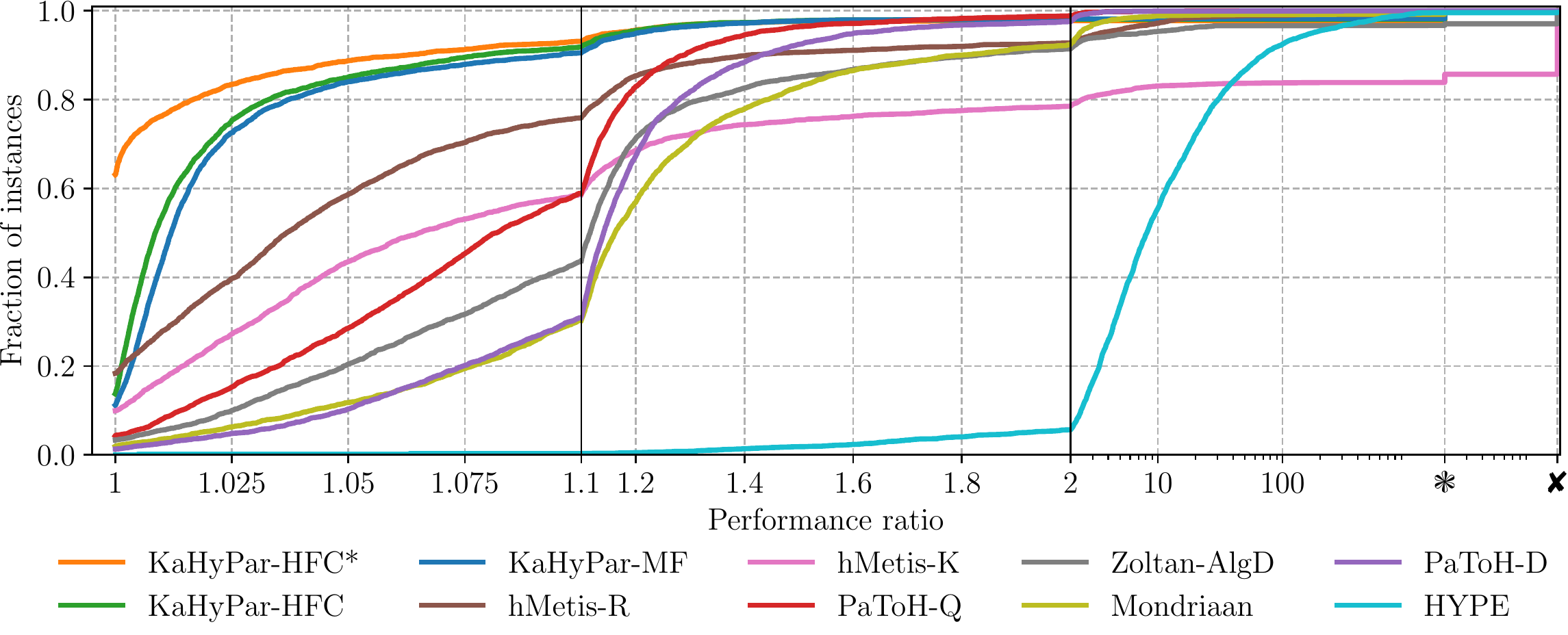}
	\caption{Performance profiles of all considered partitioners.}\label{fig:other_comparisons}
\end{figure}

\begin{figure}[tb]
	\begin{center}
		\includegraphics[width=.8\linewidth]{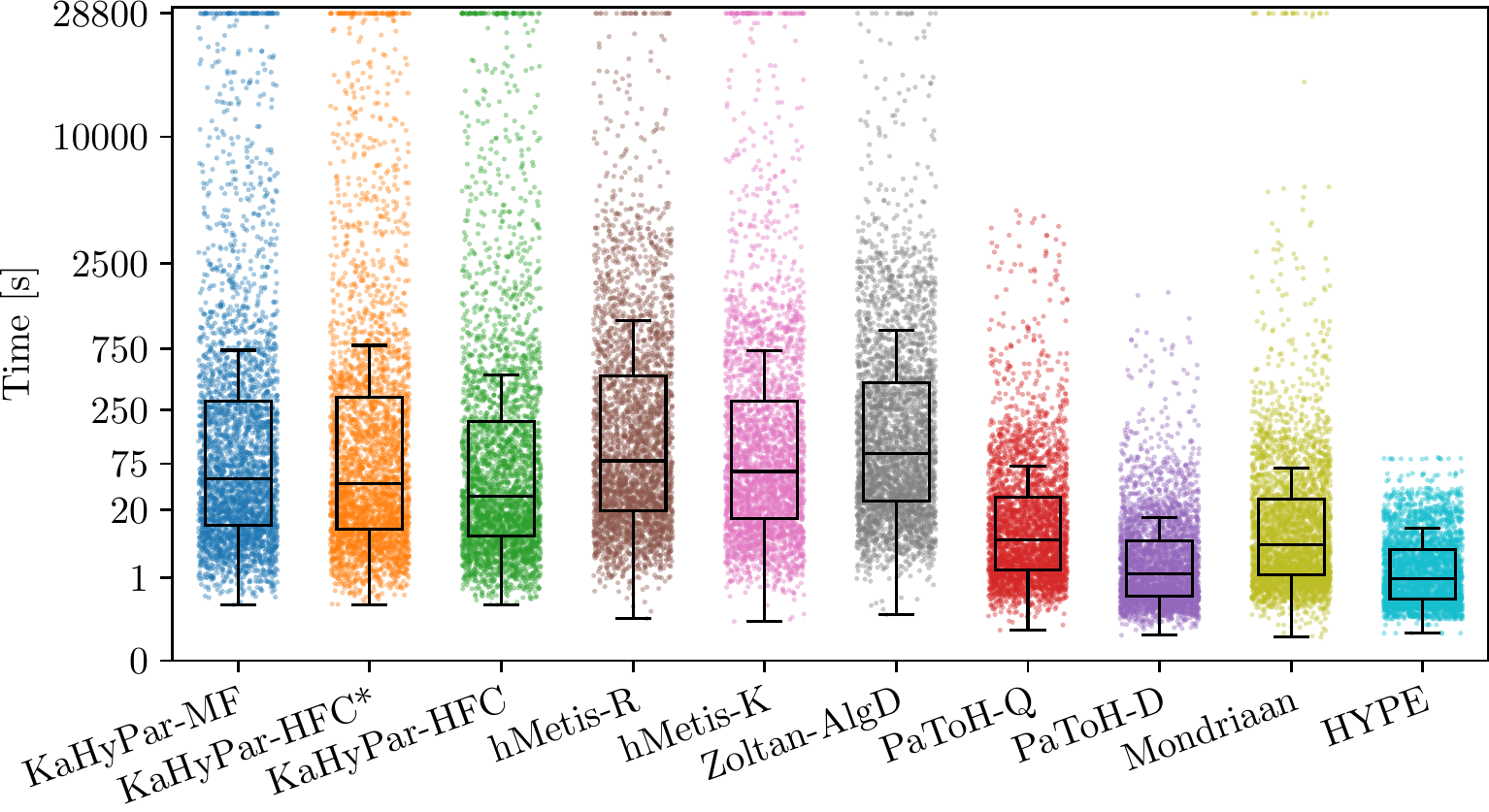}
		\caption{Box and scatter plots of arithmetic mean running times per instance.}\label{fig:running_times}
	\end{center}
\end{figure}

We now compare the three KaHyPar variants with the other state-of-the-art algorithms.
Figure~\ref{fig:other_comparisons} shows that KaHyPar-HFC* outperforms all competing algorithms, and that hMetis-R emerges as the best competitor outside the KaHyPar variants.
KaHyPar-HFC* computes the best solutions on 63\% of all instances, KaHyPar-HFC on 14\%, and hMetis-R on 18\%, as shown by their $\rho_a(1)$ values.
Note that these values alone do not permit a ranking between the algorithms.
Both KaHyPar-MF and KaHyPar-HFC compete with KaHyPar-HFC* for the best solutions on similar instances, and thus end up with a lower $\rho_a(1)$ value.
Compared individually, KaHyPar-HFC is better than hMetis-R on 69.9\% of the instances.
Additionally, KaHyPar-HFC and KaHyPar-MF approach the profile of KaHyPar-HFC* much faster.
The KaHyPar variants are all within a factor of 1.1 of the best solution on over 90\% of the instances, and within 1.4 on over 97\%, whereas hMetis-R achieves 76\% and 90\%.
PaToH-Q and PaToH-D solve more instances than hMetis-R within factors of roughly 1.2 and 1.4, and more instances than hMetis-K within 1.1 and 1.2.
Mondriaan is similar to PaToH-D and Zoltan-AlgD settles between PaToH-D and PaToH-Q.
The only non-multilevel algorithm HYPE is considerably worse, with only 5.7\% of solutions within a factor 2 of the best.

Figure~\ref{fig:running_times} shows running times for each instance.
We categorize the algorithms into two groups.
Algorithms in the first group, consisting of KaHyPar, hMetis and Zoltan-AlgD, invest substantial running time to aim for high-quality solutions.
On the other hand, PaToH, Mondriaan and HYPE aim for fast running time and reasonable solution quality.
The results show that while PaToH gives the best time-quality trade-off, KaHyPar-HFC* is the best algorithm for high-quality solutions, whereas KaHyPar-HFC offers a better time-quality trade-off than other algorithms from the first group.

In Table~\ref{table:additional_stats} in Appendix~\ref{sec:additional_stats}, we report aggregate running times, timeouts and imbalanced solutions.
Figure~\ref{fig:experimental:performance_profile_different_k} in Appendix~\ref{appendix:k_2} presents separate performance profiles for the different values of $k$.
These show that the performance difference between KaHyPar and the other algorithms increases with $k$.
This can be explained by the fact that all other partitioners except hMetis-K use recursive bisection.
Furthermore, Figure~\ref{fig:experimental:performance_profile_instance_classes} in Appendix~\ref{appendix:instance_classes} shows performance profiles for each instance class of the benchmark set.
Especially on dual SAT instances that have many large hyperedges KaHyPar-HFC* improves upon KaHyPar-MF.
Finally, Figure~\ref{fig:best_in_family} in Appendix~\ref{appendix:best_in_family} presents performance profiles comparing KaHyPar-HFC* and KaHyPar-HFC with the best algorithm from each family of partitioning algorithms.

\section{Conclusion and Future Work}\label{sec:conclusion}

We leverage the powerful HyperFlowCutter refinement algorithm in the multilevel setting for $k$-way partitioning by integrating it into KaHyPar.
For this, we extend unweighted HyperFlowCutter to weighted hypergraphs by adapting its balancing heuristics and presenting an approach to compute flows directly on weighted hypergraphs.
Furthermore, we propose a distance-based piercing heuristic and use the existing avoid-augmenting-paths piercing heuristic to find partitions with small imbalance.

The most pressing area of research is to reduce the running time when using large flow hypergraphs, \eg, by further pruning of scheduled block pairs or more advanced flow algorithms like (E)IBFS~\cite{ibfs, eibfs}.
Furthermore, the impact of HFC refinement with small flow hypergraphs on fast and medium-quality partitioners such as PaToH could be a promising direction, since previous work on bipartitioning~\cite{ghw-efbhba-esa19} already gave promising results.

\bibliography{references}

\newpage
\appendix

\section{Component Evaluation}\label{sec:component_eval}

\begin{figure}[h!]
	\includegraphics[width=.5\linewidth]{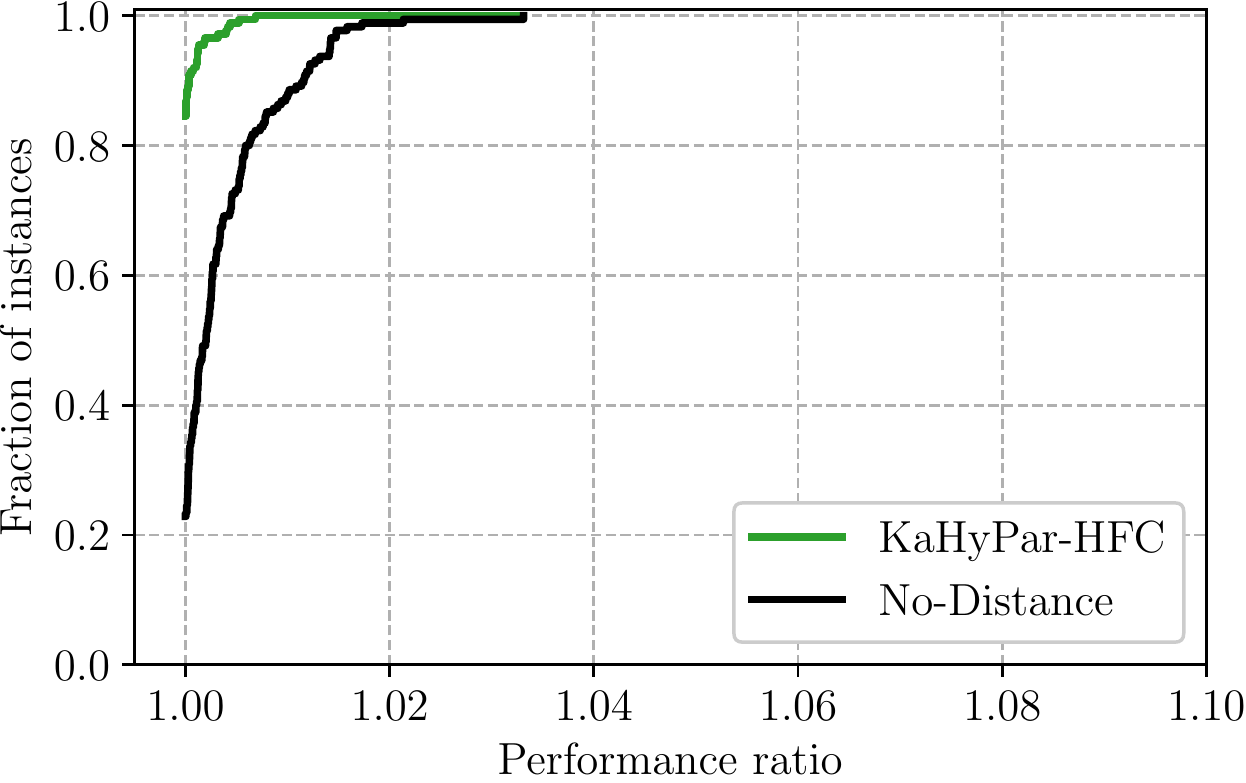}
	\includegraphics[width=.5\linewidth]{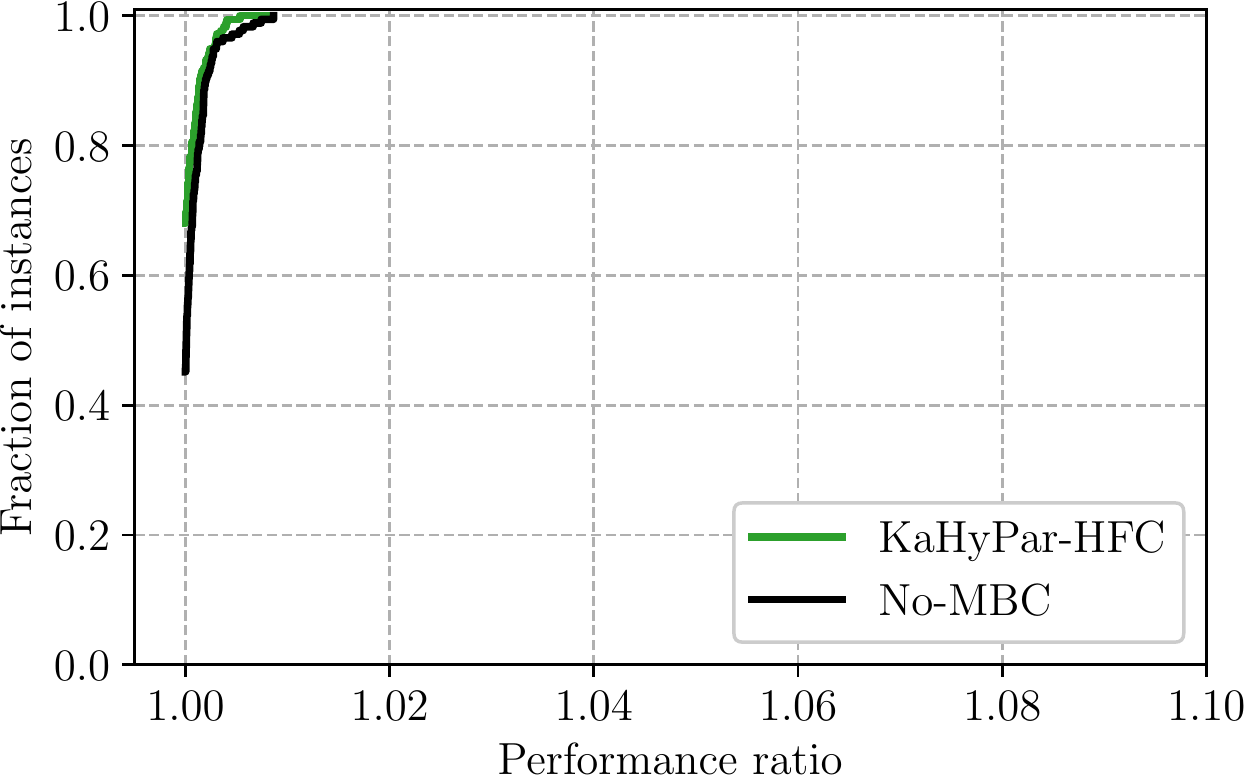}
	\includegraphics[width=.5\linewidth]{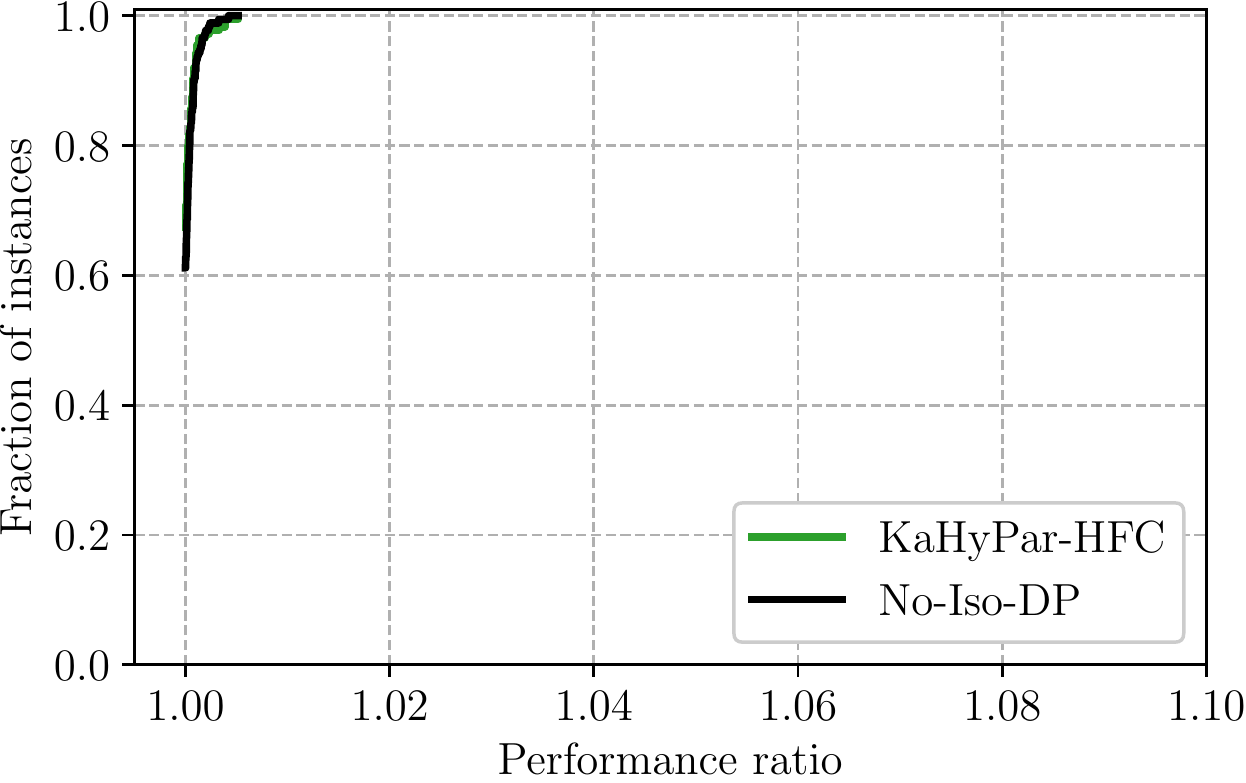}
	\caption{Assessing the impact of the optimizations by disabling each component separately.}
	\label{fig:component_eval}
\end{figure}

We assess the impact of the additional algorithmic components (DP for isolated vertices, distance-based piercing, keep piercing/most balanced cuts) by comparing KaHyPar-HFC with a KaHyPar-HFC version where one component is disabled (No-Iso-DP, No-Distance, No-MBC).
The evaluation is performed on the parameter configuration benchmark set (set D) of KaHyPar-MF~\cite{kahypar-mf-19} containing 25 \emph{smaller} hypergraphs.
Each hypergraph is again partitioned into $k \in \{2,4,8,16,32,64,128\}$ blocks. Thus, in total the experiments include 175 instances.
The results in Figure~\ref{fig:component_eval} are quite surprising.
The distance-based piercing heuristic is quite important for the solution quality of KaHyPar-HFC and the most balanced cuts/keep piercing heuristic helps a little bit.
Surprisingly, the DP for isolated vertices had little impact on solution quality.
We use it nonetheless, because it was vital for ReBaHFC, it did not impair solution quality, and because it might work better on larger instances.

\clearpage

\section{The Benchmark Set}\label{appendix:instances}

\begin{table}[h!]
\caption{Percentiles of number of vertices, hyperedges and pins in the different instance classes of the benchmark set.}
\label{table:instance_stats}
\begin{tabular}{l rrrrrrrr}
	\toprule
 & 	{} &    min &    5\% &    25\% &    50\% &    75\% &    95\% &   max \\
	\midrule
	
	\multirow{3}{*}{\begin{sideways}SPM \end{sideways}}
	& $|V|$         & 10.0K & 11.1K &  19.1K &  45.2K & 117.6K & 914.6K &  9.8M \\
	& $|E|$         & 163.0 & 10.1K &  17.2K &  40.1K & 119.9K & 988.5K &  6.9M \\
	& $|P|$         &  6.3K & 56.3K & 232.0K & 776.5K &   2.4M &  17.1M & 96.8M \\

	\midrule

	\multirow{3}{*}{\begin{sideways}Dual \end{sideways}}
	& $|V|$         & 28.8K &  62.6K & 143.4K & 771.5K &   2.0M &  6.4M & 13.4M \\
	& $|E|$         &  7.5K &   8.5K &  32.9K & 119.1K & 326.3K &  1.5M &  1.6M \\
	& $|P|$         & 76.3K & 218.6K & 336.8K &   2.3M &   6.4M & 16.0M & 39.2M \\

	\midrule

	\multirow{3}{*}{\begin{sideways}Primal \end{sideways}}
	& $|V|$         &  7.5K &   8.5K &  32.9K & 119.1K & 326.3K &  1.5M &  1.6M \\
	& $|E|$         & 28.8K &  62.6K & 143.4K & 771.5K &   2.0M &  6.4M & 13.4M \\
	& $|P|$         & 76.3K & 218.6K & 336.8K &   2.3M &   6.4M & 16.0M & 39.2M \\

	\midrule

	\multirow{3}{*}{\begin{sideways}Literal \end{sideways}}
	& $|V|$         & 15.0K &  16.9K &  65.7K & 238.2K & 652.7K &  2.9M &  3.2M \\
	& $|E|$         & 28.8K &  62.6K & 143.4K & 771.5K &   2.0M &  6.4M & 13.4M \\
	& $|P|$         & 76.3K & 218.6K & 336.8K &   2.3M &   6.4M & 16.0M & 39.2M \\

	\midrule

	\multirow{3}{*}{\begin{sideways}DAC \end{sideways}}
	& $|V|$         & 522.5K & 571.2K & 734.8K & 935.2K & 1.0M & 1.3M & 1.4M \\
	& $|E|$         & 511.7K & 560.3K & 731.5K & 916.9K & 1.0M & 1.3M & 1.3M \\
	& $|P|$         &   1.7M &   1.9M &   2.4M &   3.1M & 3.3M & 4.9M & 4.9M \\

	\midrule

	\multirow{3}{*}{\begin{sideways}ISPD98 \end{sideways}}
	& $|V|$         & 12.8K & 18.6K &  30.1K &  61.4K & 131.8K & 189.3K & 210.6K \\
	& $|E|$         & 14.1K & 18.8K &  32.7K &  68.0K & 139.5K & 191.8K & 201.9K \\
	& $|P|$         & 50.6K & 76.6K & 126.8K & 251.4K & 499.4K & 825.7K & 860.0K \\
	
	\bottomrule
\end{tabular}
\end{table}

\FloatBarrier
\clearpage

\section{Aggregate Running Times and Failed Runs}\label{sec:additional_stats}

On one instance, HYPE output an infeasible objective value.
Mondriaan reported an error on 4 instances.
Regarding infeasible solutions, hMetis-K computes imbalanced partitions on 484 instances.
To the best of our understanding, the hMetis-K paper~\cite{kk-mkwhp-99} does not mention a constraint or penalty on maximum vertex weights during coarsening.
Heavy vertices make it difficult for initial partitioning to find balanced partitions, which could be a possible explanation for these effects.

\begin{table}[h!]
	\caption{Overview of geometric mean running times and number of instances with timeouts, errors, or imbalanced partitions.}
	\label{table:additional_stats}
	\begin{tabular}{l *{5}r}
	\toprule	
		& KaHyPar-HFC* & KaHyPar-HFC & KaHyPar-MF & hMetis-R & hMetis-K  \\
		\midrule
		Gmean time (s) & 62.49 & 44.84 & 67.07 & 96.55 & 73.65 \\
		Timeouts & 75 & 66 & 63 & 63 & 27  \\
		Imbalanced & 0 & 0 & 0 & 0 & 484 \\
		Error & 0 & 0 & 0 & 0 & 0 \\
		\\
		
		& PaToH-Q & PaToH-D & Zoltan-AlgD & Mondriaan & HYPE \\
		\midrule
		Gmean time (s) & 7.48 & 1.47 & 107.60 & 6.44 & 1.03 \\
		Timeouts & 0 & 0 & 12 & 19 & 0  \\
		Imbalanced & 0 & 0 & 99 & 3 & 0 \\
        Error & 0 & 0 & 0 & 4 & 1 \\
        \bottomrule
	\end{tabular}
\end{table}

In a rerun of one seed of the experiments, where we disabled some expensive timing measurements, KaHyPar-HFC* only times out on 60 instances and KaHyPar-HFC on 57.
These numbers are more consistent with the reported running times than the timeouts reported in Table~\ref{table:additional_stats}.
On the instances that still time out, initial partitioning often dominates the running time.
This can be solved by implementing techniques to sparsify the coarsest hypergraph.

\clearpage
\section{Comparing All Partitioners For Different Values of $k$}\label{appendix:k_2}
\vspace{-.05cm}
\begin{figure}[h!]
	\begin{subfigure}{.5\linewidth}
		\includegraphics[width=\linewidth]{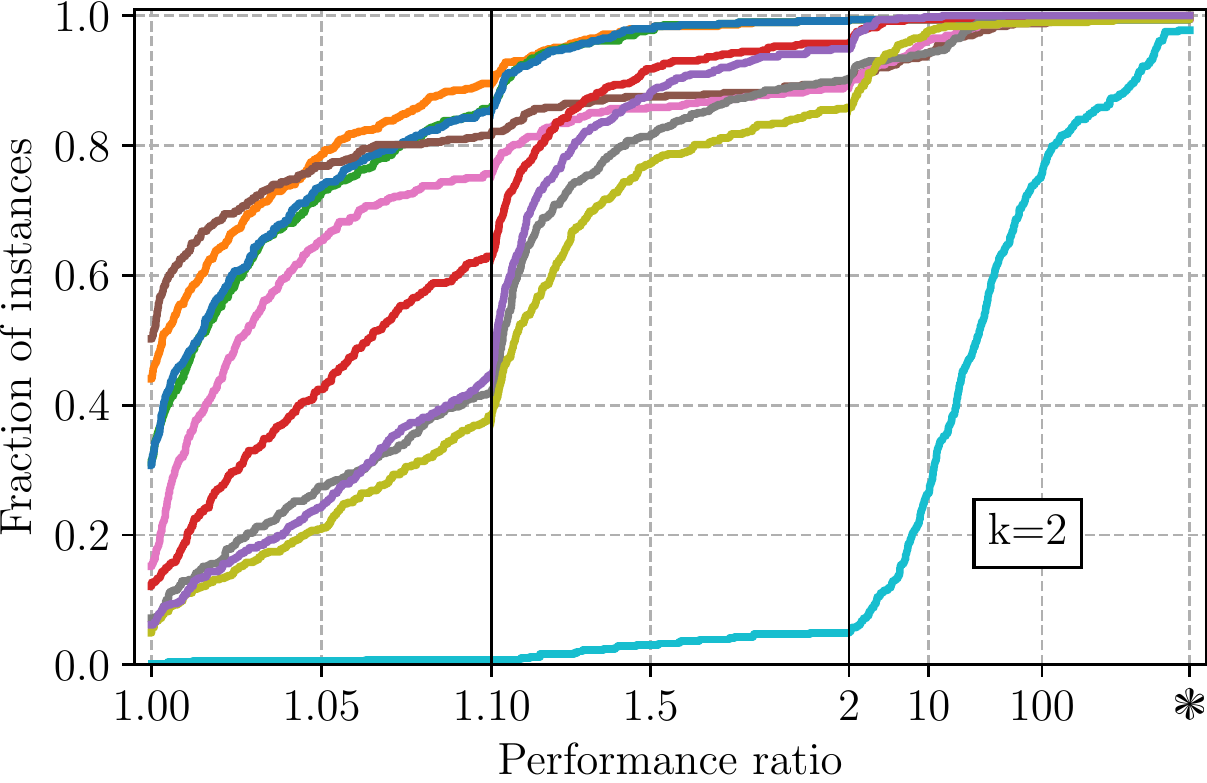}
	\end{subfigure}
	\hfill
		\begin{subfigure}{.5\linewidth}
		\includegraphics[width=\linewidth]{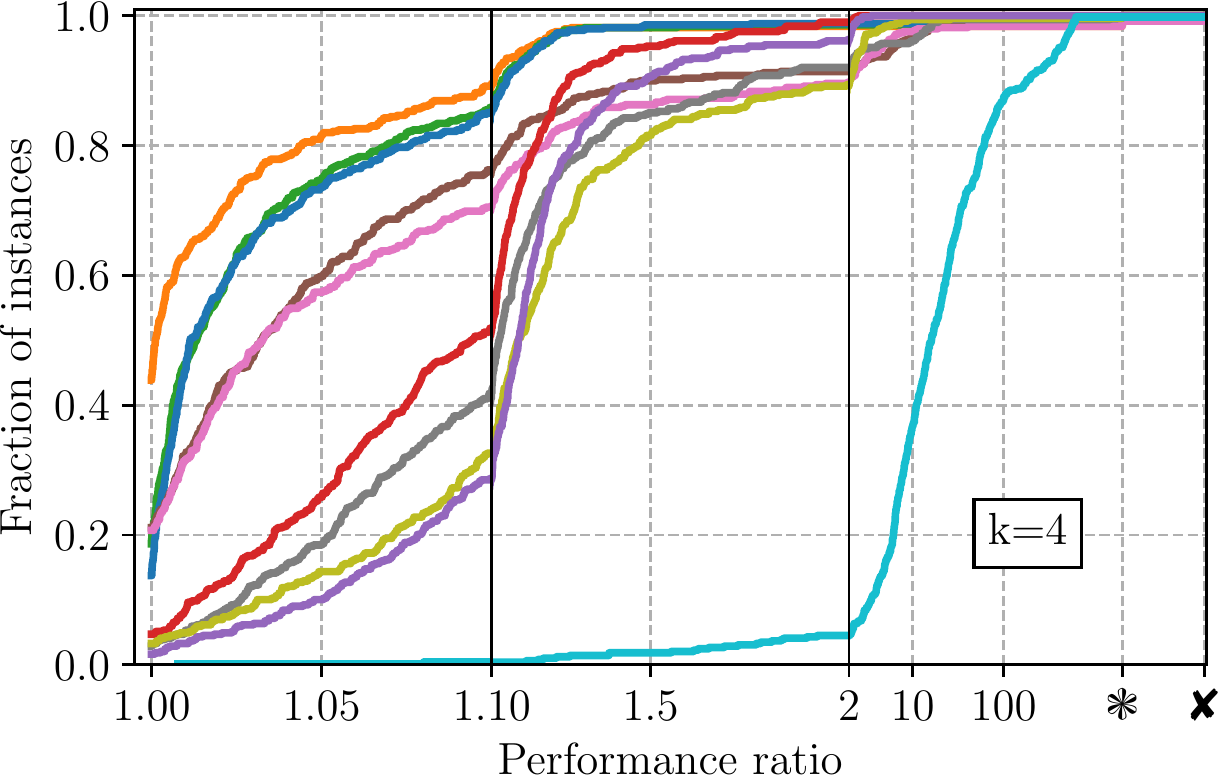}
	\end{subfigure}
	\\

	\begin{subfigure}{.5\linewidth}
	\includegraphics[width=\linewidth]{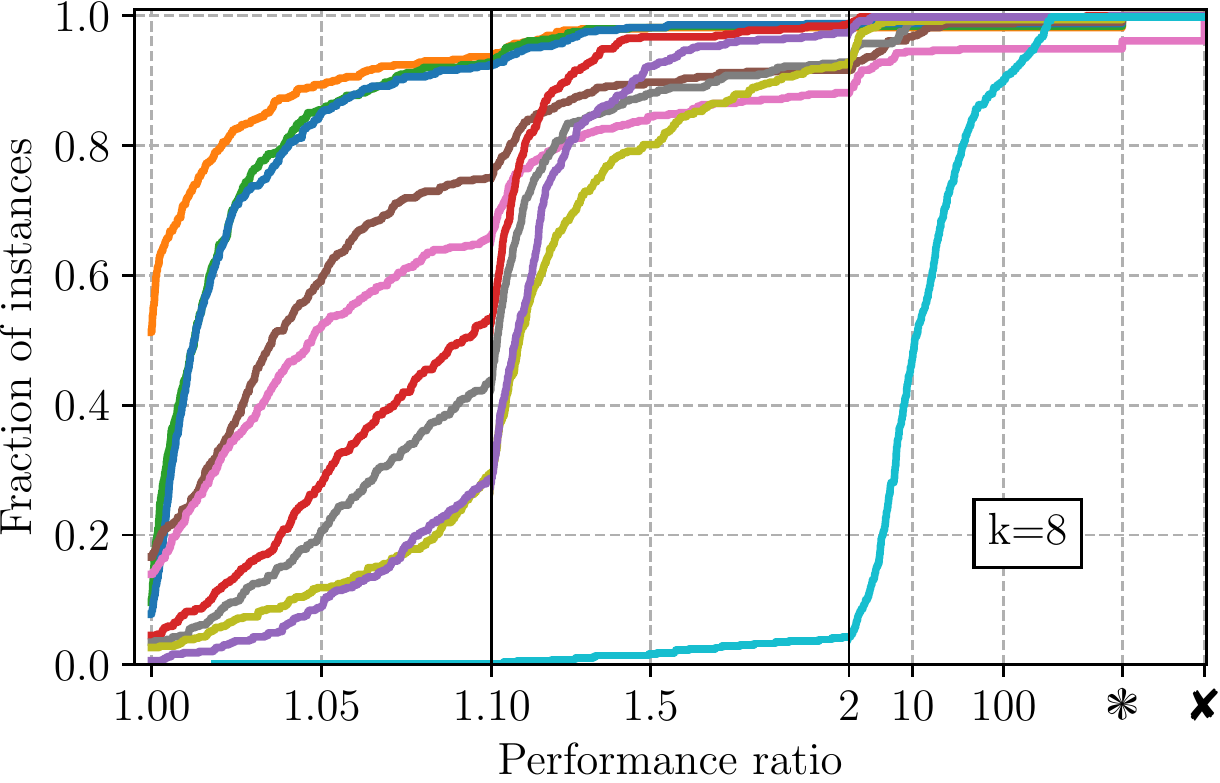}
	\end{subfigure}
	\hfill
	\begin{subfigure}{.5\linewidth}
		\includegraphics[width=\linewidth]{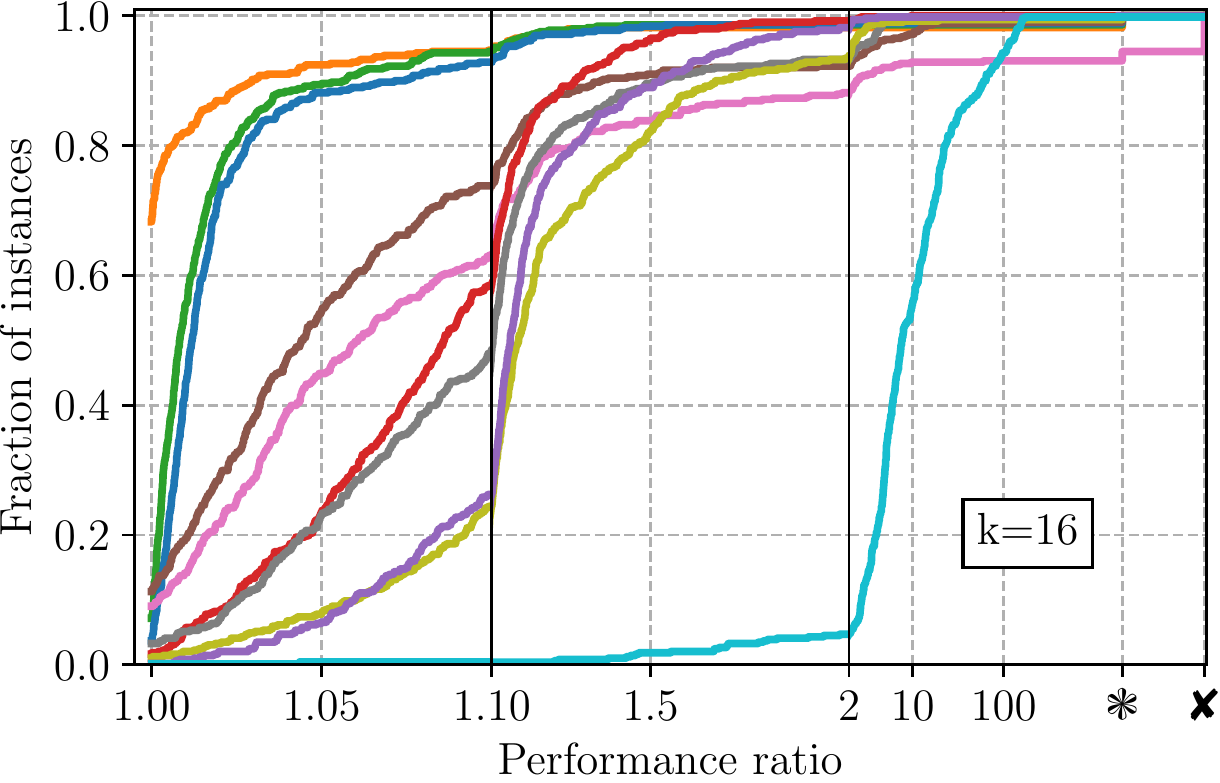}
	\end{subfigure}
	\\
	
	\begin{subfigure}{.5\linewidth}
	\includegraphics[width=\linewidth]{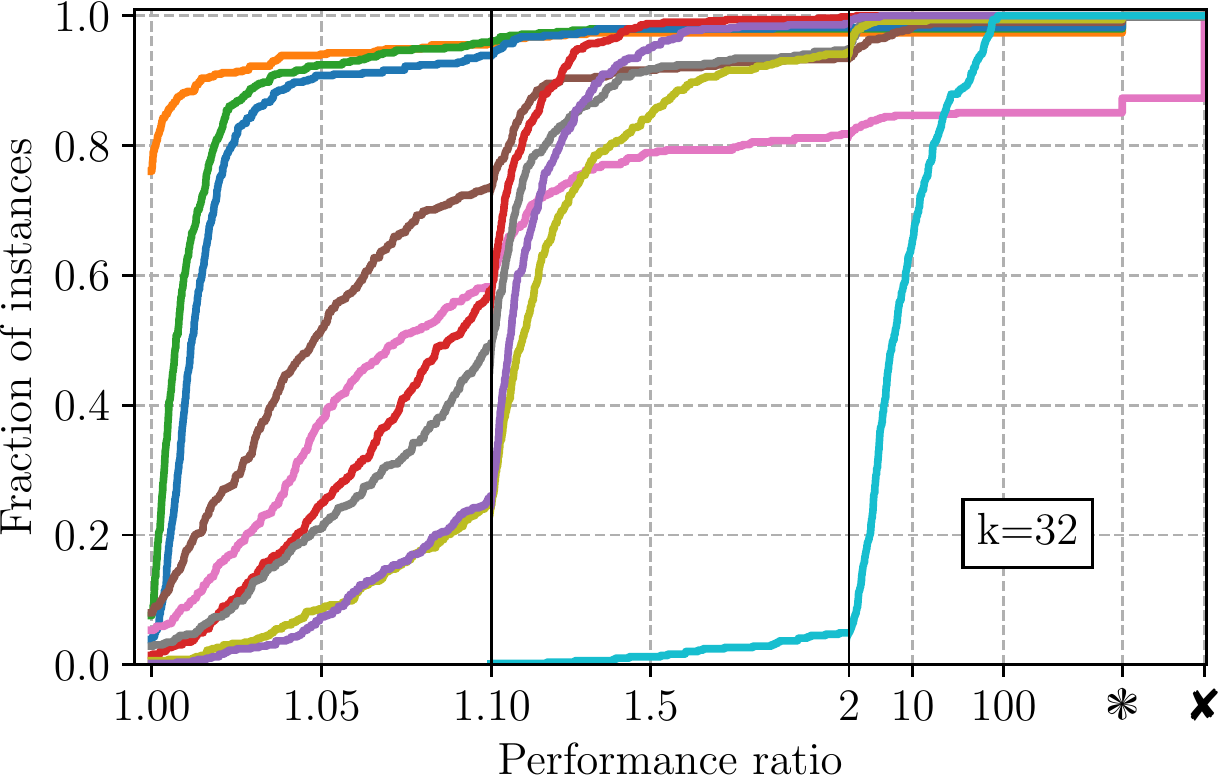}
	\end{subfigure}
	\hfill
	\begin{subfigure}{.5\linewidth}
		\includegraphics[width=\linewidth]{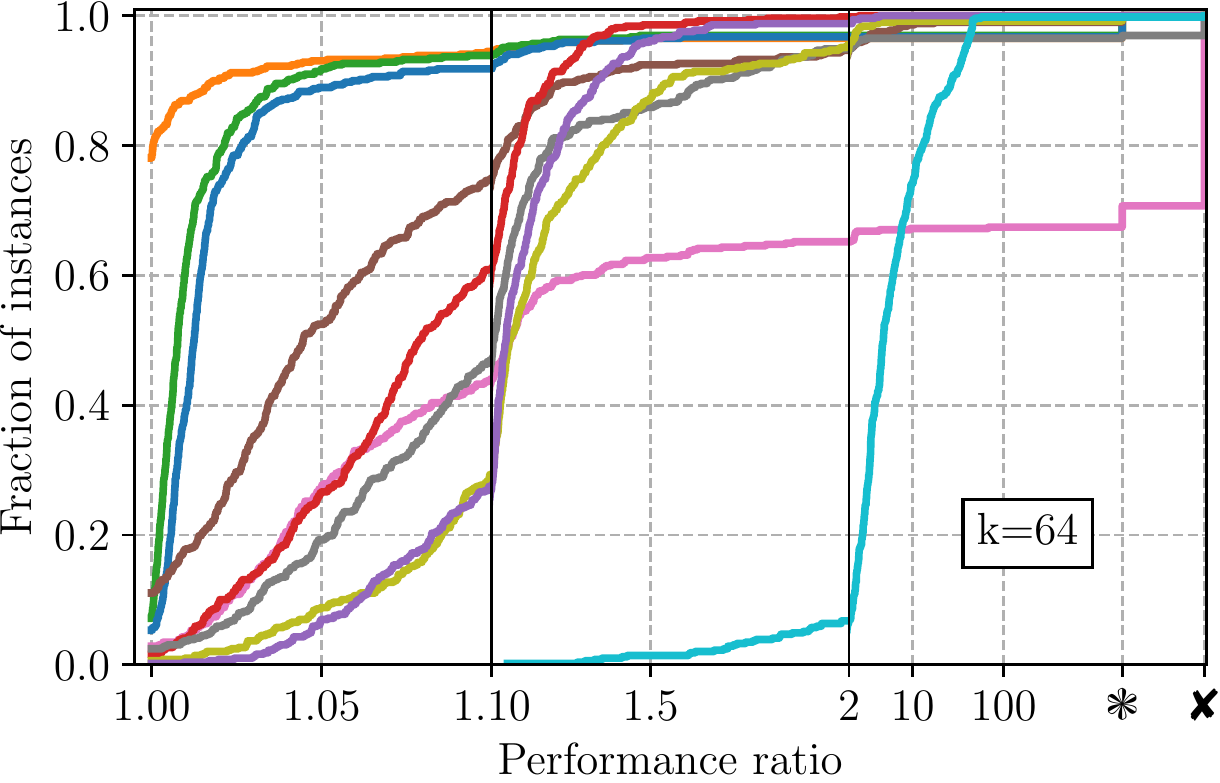}
	\end{subfigure}
	\\
	
	\begin{subfigure}{.5\linewidth}
	\includegraphics[width=\linewidth]{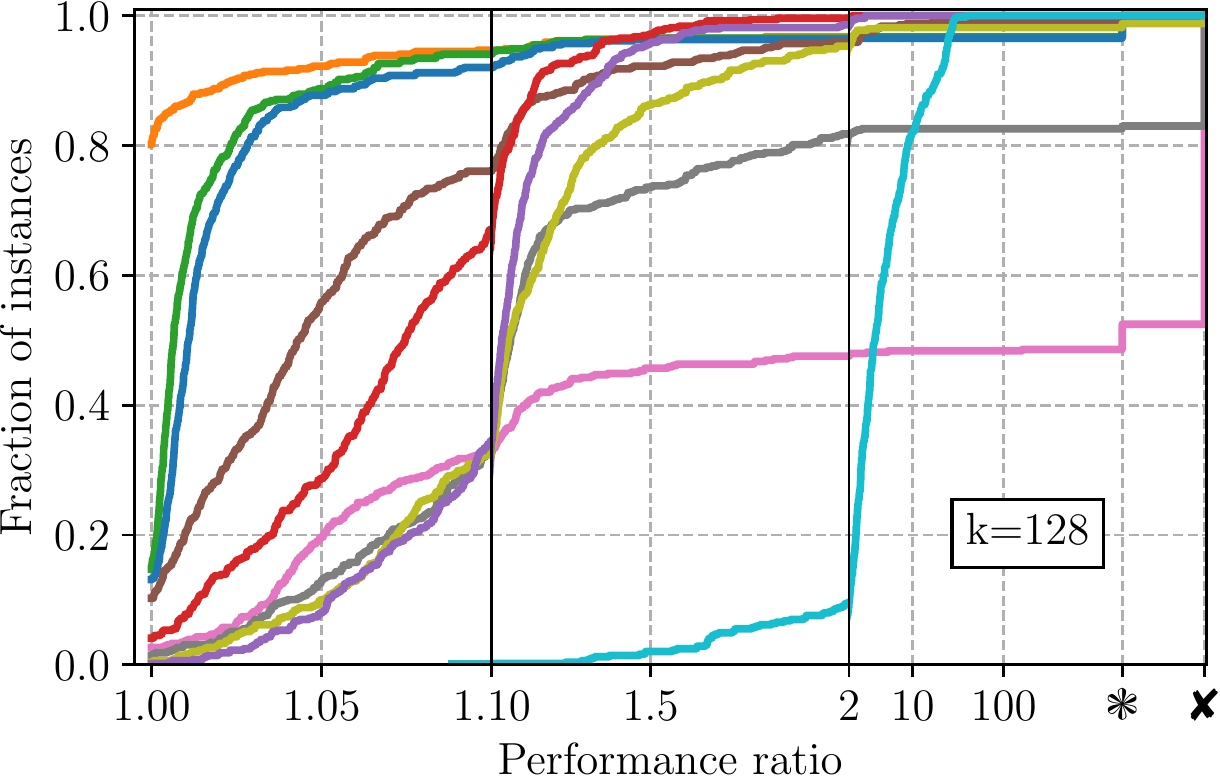}
	\end{subfigure}
	\hfill
	\begin{subfigure}{.5\linewidth}
		\begin{center}
		\includegraphics[width=0.8\linewidth]{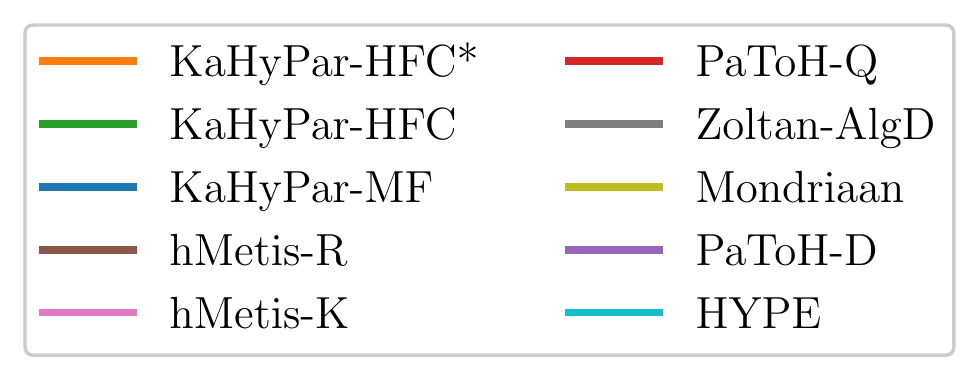}
		\end{center}
	\end{subfigure}
	\\
	
	\caption{Performance profiles for different values of $k$.}
	\label{fig:experimental:performance_profile_different_k}
\end{figure}

\FloatBarrier
\clearpage

\section{Comparing All Partitioners For Different Instance Classes}\label{appendix:instance_classes}
\vspace{-.05cm}
\begin{figure}[h!]
	\begin{subfigure}{.5\linewidth}
		\includegraphics[width=\linewidth]{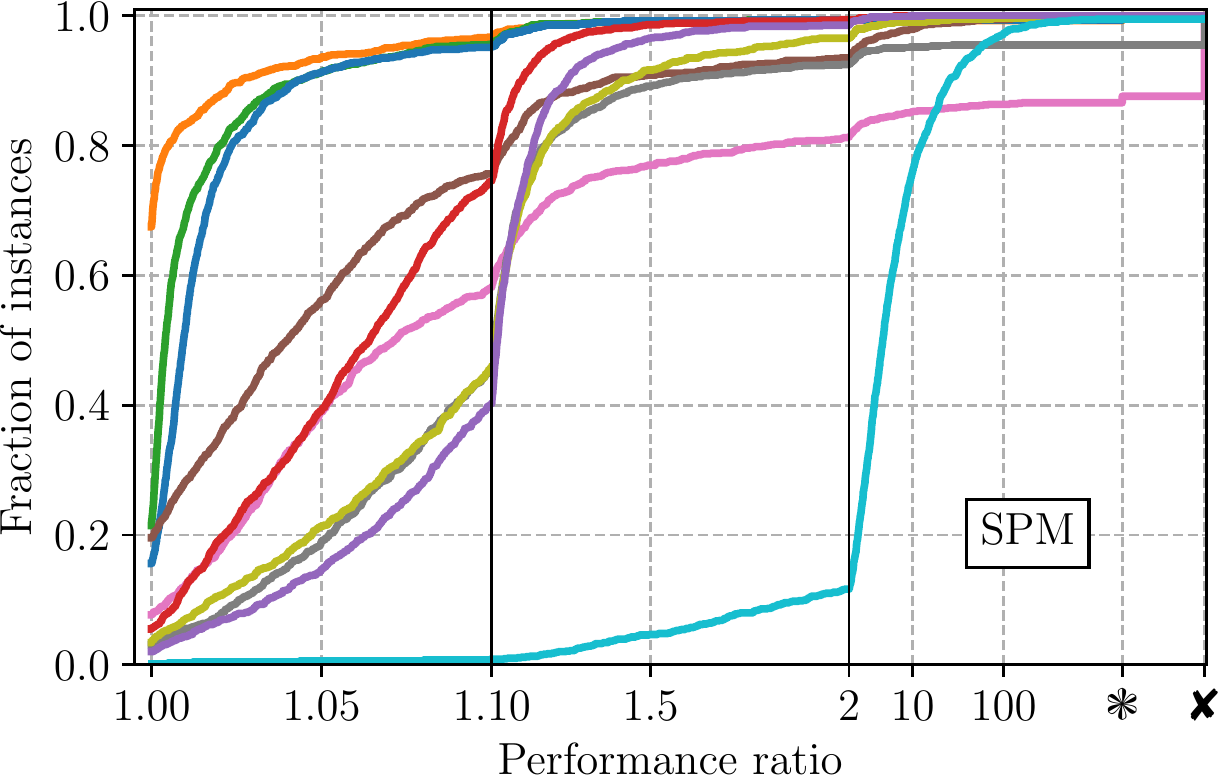}
	\end{subfigure}
	\hfill
	\begin{subfigure}{.5\linewidth}
		\includegraphics[width=\linewidth]{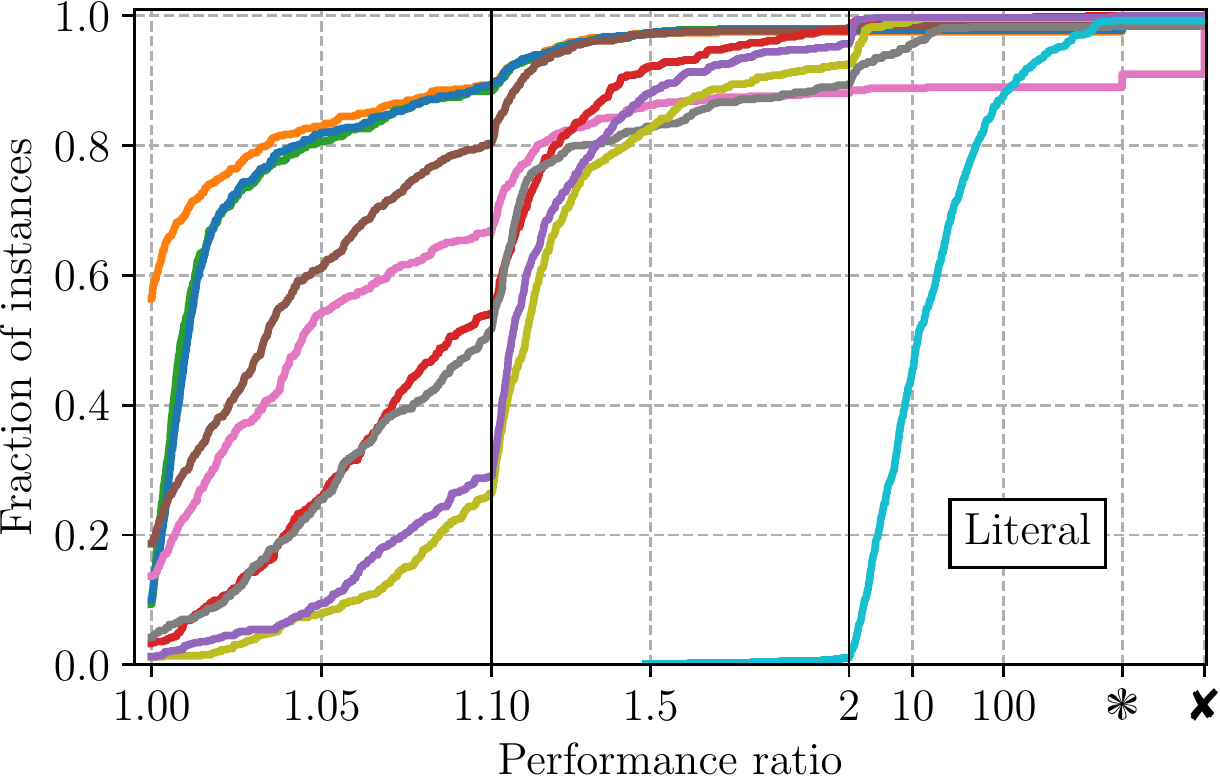}
	\end{subfigure}
	\\
	
	\begin{subfigure}{.5\linewidth}
		\includegraphics[width=\linewidth]{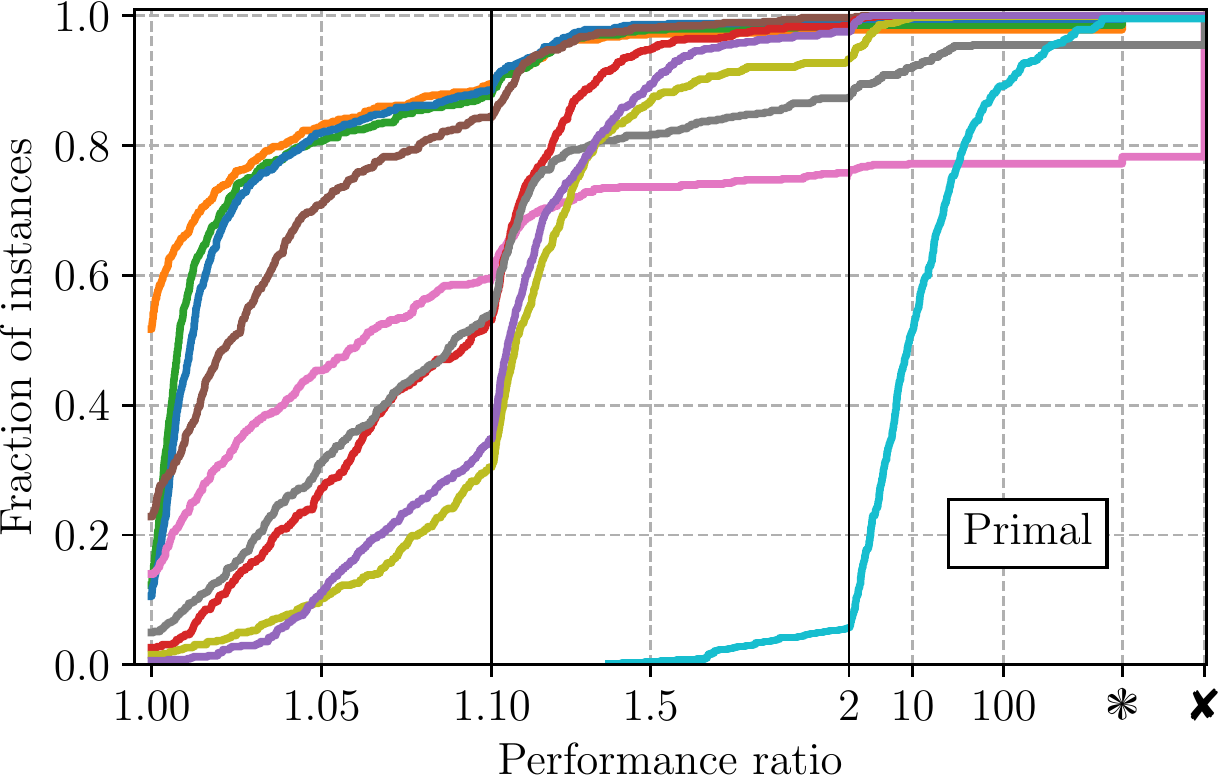}
	\end{subfigure}
	\hfill
	\begin{subfigure}{.5\linewidth}
		\includegraphics[width=\linewidth]{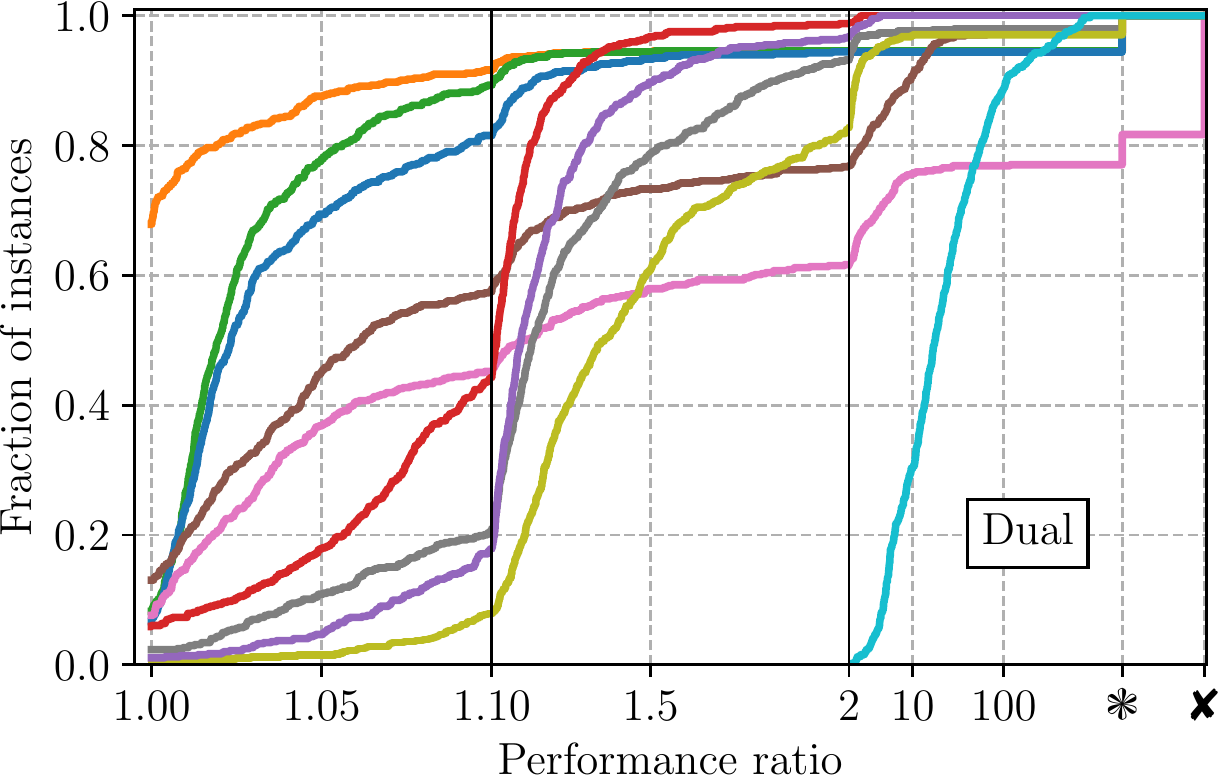}
	\end{subfigure}
	\\
	
	\begin{subfigure}{.5\linewidth}
		\includegraphics[width=\linewidth]{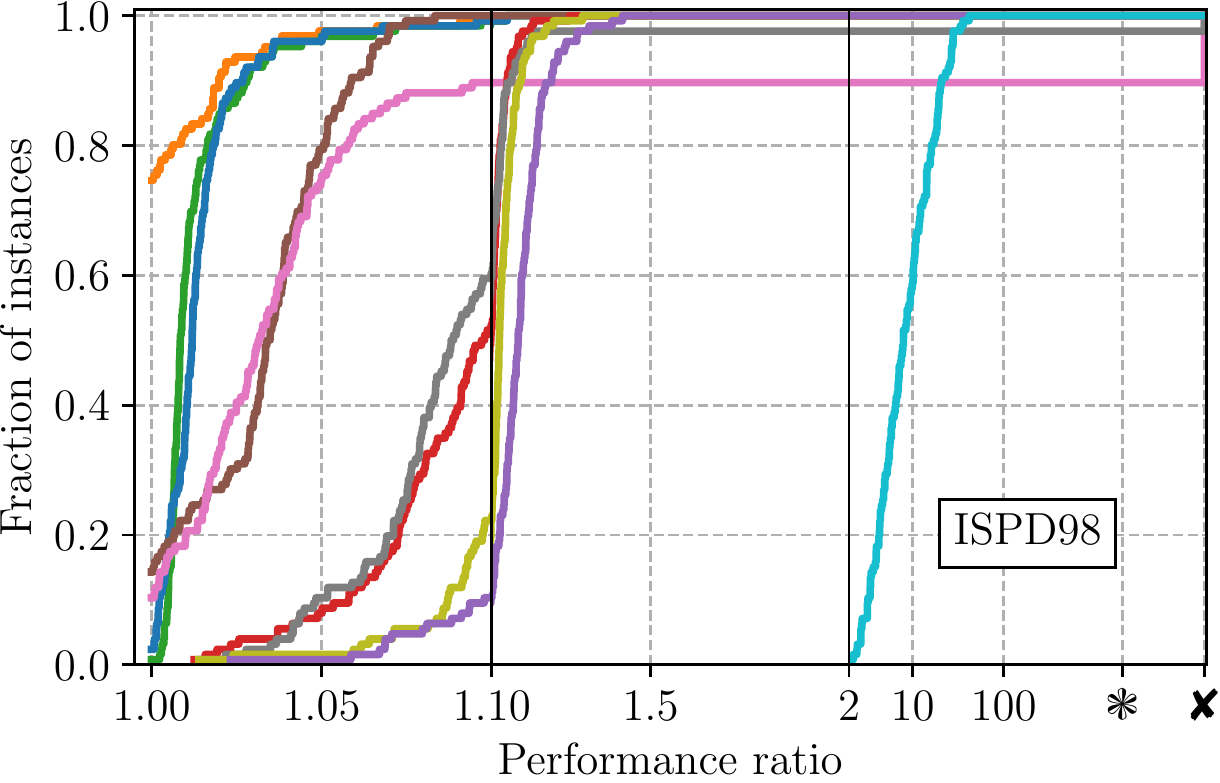}
	\end{subfigure}
	\hfill
	\begin{subfigure}{.5\linewidth}
		\includegraphics[width=\linewidth]{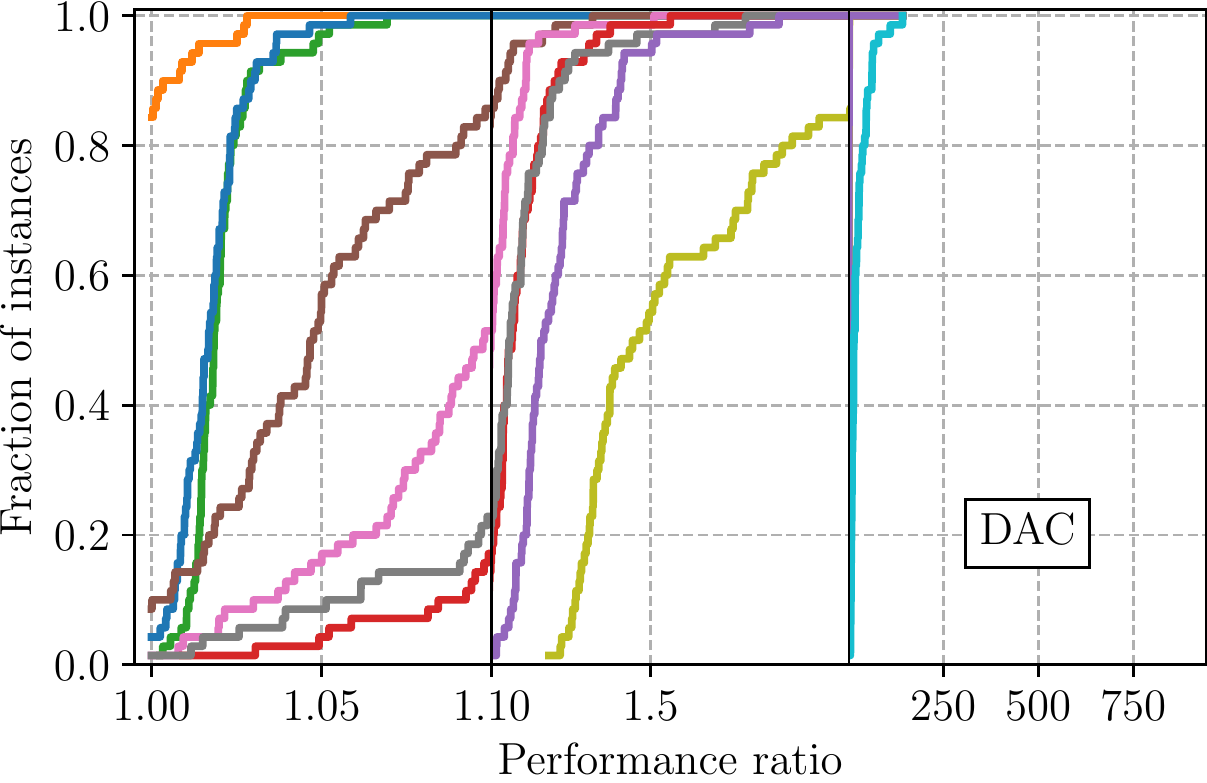}
	\end{subfigure}
	\\
	
	\begin{subfigure}{\linewidth}
		\begin{center}
		\includegraphics[width=0.4\linewidth]{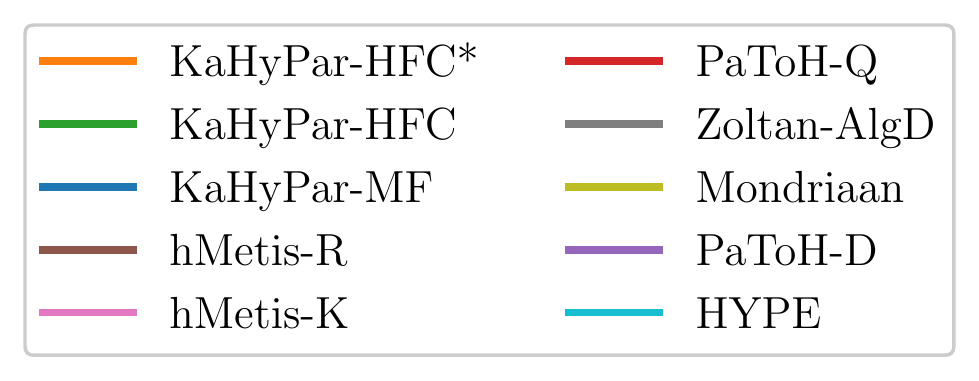}
		\end{center}
	\end{subfigure}

	\caption{Performance profiles for the different instance classes of the benchmark set.}
	\label{fig:experimental:performance_profile_instance_classes}
\end{figure}

\FloatBarrier
\clearpage

\clearpage

\section{Comparing the Best Partitioners From Each Family}\label{appendix:best_in_family}
\begin{figure}[h!]
	\includegraphics[width=.5\linewidth]{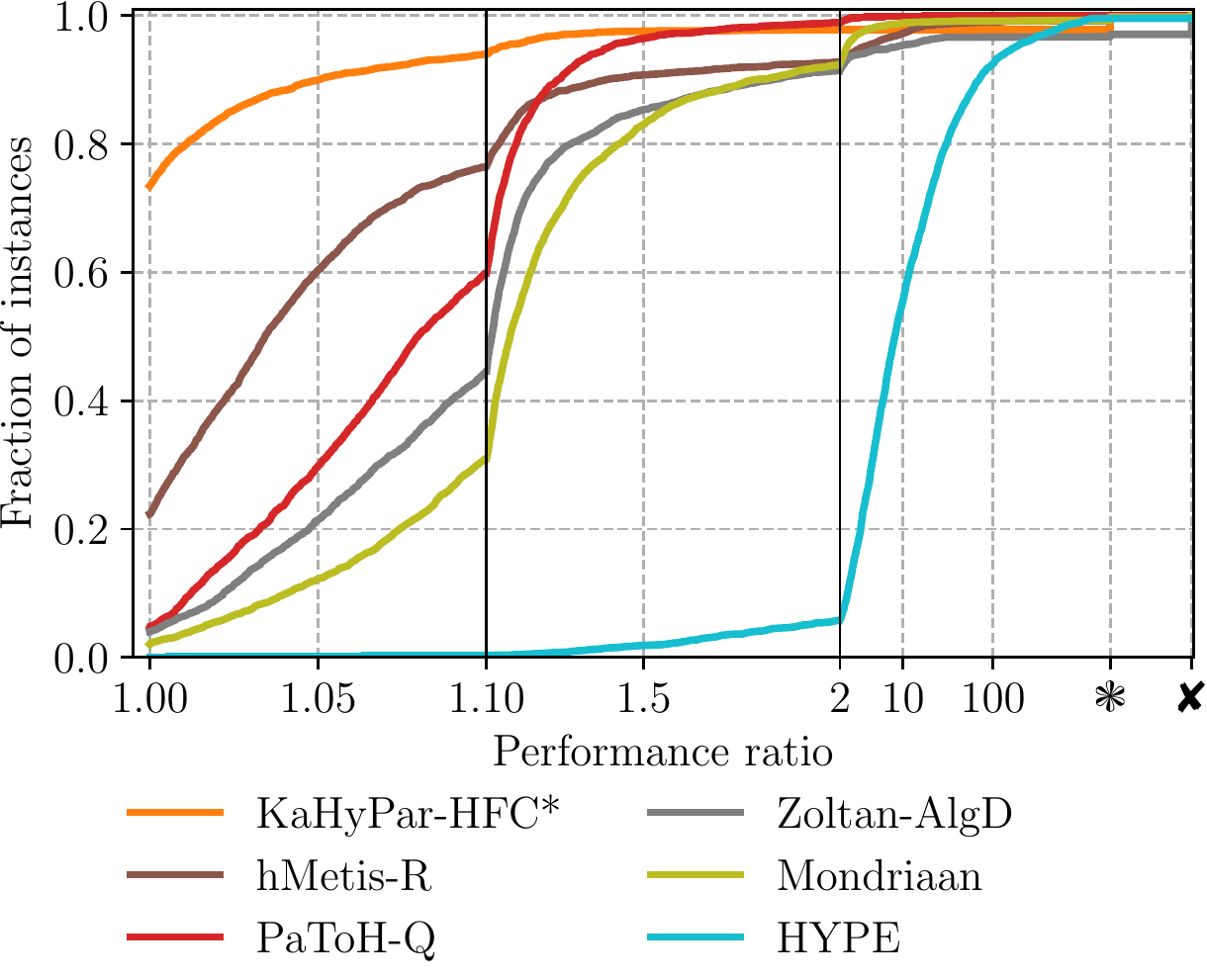}
	\includegraphics[width=.5\linewidth]{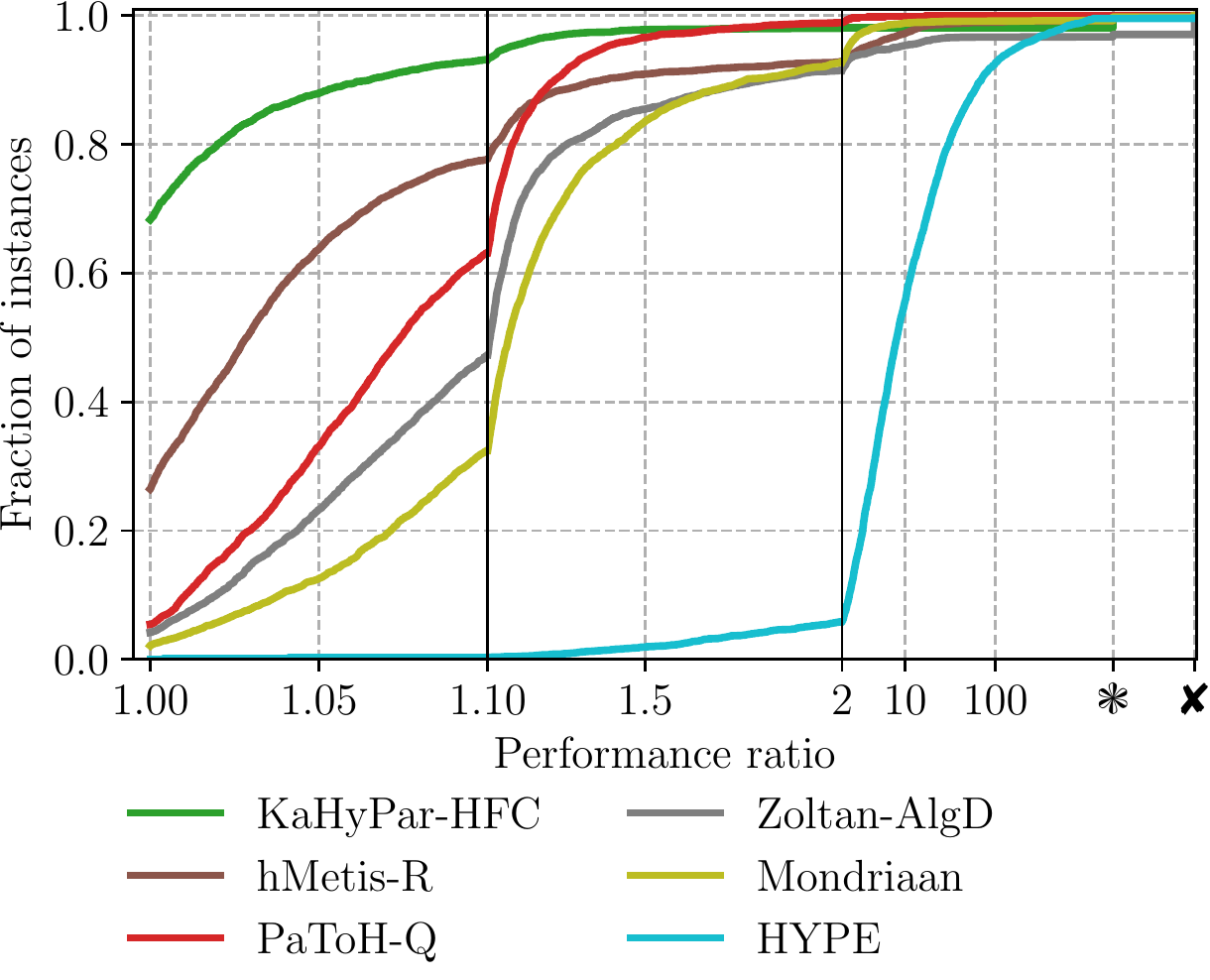}
	\caption{Comparing the best partitioner from each family, \ie, PaToH-D, hMetis-K and KaHyPar-MF were excluded, KaHyPar-HFC* and KaHyPar-HFC are considered separately.}
	\label{fig:best_in_family}
\end{figure}

\end{document}